\documentclass[12pt]{article}
\pdfoutput=1 % for arXiv
\usepackage{amsfonts,amsmath}
\usepackage[a4paper,margin=1in]{geometry}
\usepackage{hyperref}
\usepackage{graphicx}
\usepackage{verbatim}
\usepackage{color}
\usepackage[usenames,dvipsnames]{xcolor}

\newcommand{\bea}{\begin{eqnarray}}
\newcommand{\eea}{\end{eqnarray}}
\newcommand{\ee}{\end{equation}}
\newcommand{\be}{\begin{equation}}

\def\Re{{\rm Re}}
\def\Im{{\rm Im}}

\def\half{ {1\over 2}}
\def\p{\partial}

\def\pw{\partial_w}
\def\a{\alpha}
\def\b{\beta}
\def\ba{\bar \alpha}
\def\bb{\bar \beta}

\def\ep{\varepsilon}

\def\l{\lambda}
\def\o{\omega}

\def\f{\varphi}
\def\G{\Gamma}

\def\as{\alpha_s}
\def\bs{\beta_s}
\def\abs{\bar\alpha_s}
\def\bbs{\bar\beta_s}
\def\no{\nonumber}

\def\k{\kappa}

\def\oa{\omega_{\alpha}}
\def\ob{\omega_{\beta}}
\def\oab{\bar\omega_{\alpha}}
\def\obb{\bar\omega_{\beta}}
\begin{document}

\begin{titlepage}

\begin{center}

\vskip 1in

{\Large \bf Exact Half-BPS Flux Solutions in M-theory with $D(2, 1; c'; 0)^2$ Symmetry:
Local Solutions}
\vskip 0.9in

{\large John Estes$^{1 a}$, Roman Feldman$^{2 b}$, Darya Krym$^{13 c}$}

\begin{flushleft}\small

\vskip 5mm

\centerline{$^1$ Institute of Theoretical Physics }
\centerline{University of Leuven}
\centerline{Celestijnenlaan 200D B-3001 Leuven, Belgium}

\vskip 5mm

\centerline{$^2$ Depository Trust and Clearing Corp.}
\centerline{55 Water Str. NYC, NY 10041}

\vskip 5mm
\centerline{$^3$ Physics Department}
\centerline{ New York City College of Technology }
\centerline{The City University of New York}
\centerline{Brooklyn, New York 11201, USA}
\vskip 5mm

\centerline{ {\tt  ${}^a$johnaldonestes@gmail.com, ${}^b$feldman1948@gmail.com,}}
\centerline{ {\tt ${}^c$daryakrym@gmail.com}}

\end{flushleft}

\end{center}

\vskip 0.5in

\begin{abstract}
We construct the most general local solutions to 11-dimensional supergravity (or M-theory), which are invariant under the superalgebra $D(2, 1; c'; 0)\oplus D(2, 1; c'; 0)$ for all values of the parameter $c'$. The BPS constraints are reduced to a single linear PDE on a complex function $G$.  The physical fields of the solutions are determined by $c'$, a freely chosen harmonic function $h$, and the complex function $G$. $ h$ and $G$ are both functions on a 2-dimensional compact Riemannian manifold.  We obtain the expressions for the metric and the field strength in terms of $G$, $h$, and $c'$ and show that these are indeed valid solutions of the Einstein, Maxwell, and Bianchi equations.  Finally we give a construction of one parameter deformations of $AdS_7 \times S^4$ and $AdS_4 \times S^7$ as a function of $c'$.
\end{abstract}

\end{titlepage}

\tableofcontents

\setcounter{equation}{0}
\setcounter{footnote}{0}

%%%%%%%%%%%%%%%%%%%%%%%%%%%%%%%%%%%%%%%%%%%%%%%
%%%%%%%%%%%%%%%%%%%%%%%%%%%%%%%%%%%%%%%%%%%%%%%
\section{Introduction}
\setcounter{equation}{0}
%%%%%%%%%%%%%%%%%%%%%%%%%%%%%%%%%%%%%%%%%%%%%%%
%%%%%%%%%%%%%%%%%%%%%%%%%%%%%%%%%%%%%%%%%%%%%%%
We construct the exact local half-BPS flux solutions of 11-dimensional supergravity which are invariant under the superalgebra
$D(2, 1; c'; 0)\oplus D(2, 1; c'; 0)$.  A noteworthy feature of these solutions is that the $D(2, 1; c'; 0)$ superalgebra is the unique simple superalgebra with a continuous parameter, $c'$.  Note that $c'$ determines the fermionic generators but does not affect the bosonic generators of $D(2, 1; c'; 0)$ \cite{D'Hoker:2008ix}.  The existence of this parameter makes the family of solutions particularly rich and opens the door to finding interpolating families of solutions as a function of $c'$.

We reduce the BPS constraints to a single linear PDE for a complex function $G$, $2 \pw G=(G+\bar G)\pw\ln h $, where $h$ is a freely chosen harmonic function, while $h$ and $G$ are both functions on a 2-dimensional Riemannian manifold (possibly with boundary).  In \cite{D'Hoker:2008wc}, it was shown that the BPS constraints reduce to exactly this equation for three special values of $c'$, $c'\in\{1,-1/2,2\}$.  However, it was not known how to proceed with the reduction for arbitrary values of $c'$.  In this work we use new methods to show that the BPS constraints reduce to $2 \pw G=(G+\bar G)\pw\ln h $ for all values of $c'$.  Although this equation is independent of $c'$,
the physical fields of our solutions, i.e. the metric and the field strength components are determined by the choice of $c'$ in addition to the harmonic function $h$, and the complex function $G$.\footnote{The boundary and regularity conditions which determine a global solution do depend on $c'$, as does the relationship between the original spinor components in the BPS equations and the function $G$.}  We check that the Einstein, Maxwell, and Bianchi equations are satisfied for every such choice of $c'$, $h$ and $G$.  It is convenient to define a function $H$ related to $G$ by a fractional linear conformal transformation involving $c'$.  Although the PDE is linear for the function $G$, the physical fields are simpler in terms of $H$ (see sec. \ref{sectionsummary}).   More importantly, the range of $G$ is subject to a constraint, which when expressed as an equivalent constraint on $H$ takes the $c'$ independent form $|2H + i| < 1$.

Some of the original interest in these solutions was motivated by seeking solutions corresponding to intersecting M2 and M5 branes.  A lot of progress was made in finding these solutions in \cite{Lunin:2007ab,D'Hoker:2008wc}, where a general ansatz for the supergravity fields and supersymmetry parameters was proposed.  In particular, the bosonic subalgebra, $SO(2,2) \times SO(4) \times SO(4)$, which is independent of $c'$, is naturally realized on an $AdS_3 \times S^3 \times S^3$ fibration over a two-dimensional base space $\Sigma$.  The reduction of the BPS equations to 2-dimensions was carried out for general $c'$.  However, explicit solutions to the reduced BPS equations were only found for the special values of $c'\in\{1,-1/2,2\}$ and it was not  obvious that non-trivial solutions existed for general values of $c'$.

For the values $c'\in\{-1/2, -2\}$, the superalgebra $D(2, 1; c'; 0)\oplus D(2, 1; c'; 0)$ is simply $OSp(4^*|2) \oplus OSp(4^*|2)$ which is a subalgebra of $OSp(8^*|4)$.  This implies that the corresponding solutions admit solutions which asymptote to $AdS_7 \times S^4$, including $AdS_7 \times S^4$ itself \cite{D'Hoker:2008ix}.  The geometry $AdS_7 \times S^4$ is the near horizon geometry of M5 branes, which are conjectured to have a dual description in terms of a six-dimensional conformal field theory (CFT$_6$) \cite{Maldacena:1997re}.  Although the CFT$_6$ is not yet completely known, there has been work on understanding the theory in various limits (see for example: \cite{Howe:1997fb,Berkooz:1997cq,Claus:1997cq,Aharony:1997an,Douglas:2010iu,Lambert:2010iw,Kim:2011mv,Kim:2012av,Jafferis:2012iv}).

In \cite{Strominger:1995ac} (see also \cite{Ganor:1996nf,Howe:1997ue,Saemann:2010cp}), it was argued that the CFT$_6$ admits self-dual string operators, which are higher-dimensional analogues of Wilson lines in gauge theory.  In particular, there should exist self-dual string operators which preserve half of the supersymmetries, corresponding to the supergroup $OSp(4^*|2) \oplus OSp(4^*|2)$.  These solitons arise from considering M2 branes ending on M5 branes, much in the same way one obtains Wilson lines by considering fundamental strings ending on D-branes.  In \cite{D'Hoker:2008qm}, the dual supergravity solutions were found, which described general configurations of arbitrary numbers of M2 branes ending on M5 branes in the near horizon limit of the M5 branes.  Using the extended solutions presented in this paper, one might hope to generalize the solutions of \cite{D'Hoker:2008qm}, which would imply the existence of additional operators in the CFT$_6$ theory.  In particular, the self-dual strings should come with an additional parameter corresponding to the choice of $c'$.

For the value $c'=1$, the superalgebra $D(2, 1; c'; 0)\oplus D(2, 1; c'; 0)$ is $OSp(4|2,\mathbf{R}) \oplus OSp(4|2,\mathbf{R})$ which is a subalgebra of $OSp(8|4,\mathbf{R})$.  This implies that the corresponding family of solutions admits solutions which asymptote to $AdS_4 \times S^7$, including $AdS_4 \times S^7$ itself.  The geometry $AdS_4 \times S^7$ is the near horizon geometry of M2 branes.  Progress in the dual CFT description of M2 branes was initially made in \cite{Bagger:2006sk,Bagger:2007jr,Bagger:2007vi,Gustavsson:2007vu} and is commonly referred to as BLG theory. In \cite{Aharony:2008ug}, a generalization was proposed in terms of a supersymmetric Chern-Simons theory, which allows for an arbitrary number of M2 branes and is commonly referred to as ABJM theory.  The reduced superalgebra $OSp(4|2,\mathbf{R}) \oplus OSp(4|2,\mathbf{R})$, corresponds to deformations of BLG or ABJM theory by the insertion of $1+1$-dimensional interfaces or defects, which preserve half of the supersymmetries.

In \cite{D'Hoker:2009gg}, a specific solution in the $c'=1$ class was constructed which is dual to a deformation of ABJM by a dimension two operator.  This solution is reminiscent of the supersymmetric Janus solutions of IIB supergravity \cite{Bak:2003jk,Clark:2005te,D'Hoker:2006uu}, whose dual CFT description is discussed in \cite{Clark:2004sb,D'Hoker:2006uv,Gaiotto:2008sd}.  Progress in constructing the aforementioned dimension two dual operator on the CFT side was made in \cite{Honma:2008un} for the BLG theory, but an explicit construction of the interface operator in ABJM theory has not yet been given.

In \cite{D'Hoker:2009my}, it was shown that there are no other asymptotic $AdS_4 \times S^7$ solutions, besides the Janus solution, with the assumption that the two-dimensional base space, $\Sigma$, has disc topology.  The absence of such solutions could be interpreted to mean that ABJM theory does not allow for more general defect/interface theories other than Janus, however, there has been work in studying such deformations on the CFT side.  M2 branes ending on M5 branes from the M2 brane point of view was first studied in \cite{Basu:2004ed}.  Recent progress in studying such defects and interfaces has been made in \cite{Jeon:2008bx,Hanaki:2008cu,Jeon:2008zj,Ammon:2009wc,Berman:2009xd,Fujimori:2010ec,Fujita:2010pj}.  The existence of such objects in the CFT should imply the existence of dual gravitational solutions.  There are two ways to get around the null results of \cite{D'Hoker:2009my}.  First one may look for solutions where $\Sigma$ has other topologies, such as an annulus topology. Second, such interface and defect deformations might require one to consider values of $c'$ away from $c'=1$.

One of the principal motivations for finding solutions for general values of $c'$ is the possibility to find new families of supersymmetric solutions which can be viewed as deformations of the maximally symmetric solutions $AdS_4 \times S^7$ and $AdS_7 \times S^4$.  The existence of such families can imply the existence of corresponding operators in the dual CFTs and give predictions for how the operators can deform the CFT.  For example, the interpolation from $c'=1$, corresponding to $AdS_4 \times S^7$, to $c'\in\{0,\infty\}$ requires the decompactification of one of the $S^3$s  at the endpoints of the interpolation.  This implies that the Kaluza-Klein scale vanishes at this value of $c'$.  In the dual ABJM theory this would correspond to the closing of a ``dimension gap'', in the sense that one would expect towers of operators with continuous dimensions.  We present two explicit examples of such families of interpolating solutions at the end of this paper, one family contains $AdS_4 \times S^7$ and the other family contains $AdS_7 \times S^4$.

Perhaps the most interesting possibility is to look for a family of solutions which interpolates from $AdS_4 \times S^7$ to $AdS_7 \times S^4$.  These would be dual to a family of CFTs which would interpolate as a function of $c'$ from the 3-dimensional ABJM theory to the 6-dimensional CFT$_6$.  We note that any such interpolation must pass through a decompactification limit (a similar decompactification occurs in the generalized LLM solutions discussed in \cite{Colgain:2012wv}).  To support this idea, we note that in \cite{Gustavsson:2010ep,Terashima:2010ji,Lambert:2011eg}, there has been some progress on the CFT side in extracting some of the known CFT$_6$ data from massive ABJM theory \cite{Gomis:2008vc,Hosomichi:2008jb}.

The organization of the paper is as follows. In section \ref{sectionsetup}, we give the ansatz and review the results of \cite{D'Hoker:2008wc} which we use.  In section \ref{sectionreduction}, we present the reduction of the BPS equations to a single linear PDE.  In section \ref{sectionsummary}, we give a summary of the full solution, discuss some of the general features.  Section \ref{sectionsummary} is self-contained and a reader interested only in the solution and not in the solution methods can skip to this section.  In section \ref{sec:explicit} we construct one parameter deformations of both $AdS_7 \times S^4$ and $AdS_4 \times S^7$.  In the Appendix, we provide the derivations of the expressions for the metric and the field strength in terms of $h$ and $H$. We also show that the solutions of the BPS equations solve the Bianchi identities, as well as the Maxwell and Einstein equations of 11-dimensional supergravity.

\section{Setup}\label{sectionsetup}
\setcounter{equation}{0}

In this section we first give the $D(2, 1; c'; 0)\oplus D(2, 1; c'; 0)$ invariant ansatz for the 11-dimensional supergravity fields and then discuss the reduction of the corresponding 11-dimensional BPS equations to 2-dimensions.
This section is a review of results from \cite{D'Hoker:2008wc} and for a fuller understanding, we refer the reader to sections 2-4 of \cite{D'Hoker:2008wc}. The reader familiar with \cite{D'Hoker:2008wc} can skip this section.

We will look for bosonic solutions to 11-dimensional supergravity which is defined by the following action
\begin{align}
\label{acteq}
S={1\over 2 \kappa_{11}^2} \int d^{11}x \sqrt{-g} \Big(R-{1\over 48} F_{MNPQ}F^{MNPQ}\Big) -{1\over 12 \kappa_{11}^2} \int C \wedge F \wedge F
\end{align}
As usual, $\kappa_{11}$ is the 11-dimensional gravitational coupling, which is simply the 11-dimensional Newton's constant times a factor of $16\pi$, $g$ is the determinant of the metric and $R$ is its Ricci scalar, $F$ is the 4-form field-strength and $C$ is its 3-form gauge potential, $F = dC$. The equations of motion are given in section \ref{sectionbianchi}. In the above action, the fermionic field, the gravitino, has already been set to zero.  This is always a consistent choice.  However, in order for the solution to preserve supersymmetry, the variation of the gravitino with respect to some supersymmetry parameter $\ep$ must vanish. We will call such an $\ep$ an 11-dimensional Killing spinor. It is a 32 component Majorana spinor in 11-dimensions and the vanishing of the variation of the gravitino, called the BPS constraint, is given by the equation
\begin{align}
\label{fullgravitino}
\nabla_M \ep +{1\over 24^2} \Big( - \Gamma_M (\G \cdot F)
+ 3 (\G \cdot F) \G_M \Big) \ep=0
\end{align}
where $\nabla _M$ is the covariant derivative with respect to the Levi-Civita connection for the metric. $\G_M$ are the usual gamma matrices satisfying the 11-dimensional Clifford algebra $\{\G_a,\G_b\}=2\eta_{ab}I_{32}$, where $\eta_{ab}$ is the Lorentz metric in 11-dimensions and $I_{32}$ is the 32-dimensional identity matrix. $\G\cdot F$ means the contraction of a rank 4 anti-symmetric tensor of $\G$ matrices with the field strength.  In general, there is a supersymmetry for every linearly independent $\ep$ which satisfies \eqref{fullgravitino}.

We will look for bosonic solutions invariant under the $SO(2,2) \times SO(4) \times SO(4)$ isometry, as this is the maximal bosonic subalgebra of $D(2, 1; c'; 0)\oplus D(2, 1; c'; 0)$ for any value of $c'$. This leads to the following ansatz.  The metric has the form
\begin{align}
ds^2= f_1^2 \;ds_{AdS_3}^2+ f_2^2 \;ds_{S^3_2}^2
+ f_3^2 \; ds_{S^3_3}^2+ ds_{\Sigma}^2
\end{align}
where $ds_{AdS_3}^2$ is the unit radius metric on 1+2 dimensional anti-de-Sitter space while $ds_{S^3_2}^2$ and $ds_{S^3_3}^2$ are unit radius metrics on the 3-sphere. The metric component $ds^2 _\Sigma$ is a metric on a 2-dimensional Riemann surface $\Sigma$.

It is convenient to introduce a frame $e^A \equiv dx^M \, e_M {}^A$, with $A=0,1, \cdots , 9, \natural$,  where $\natural$ denotes the 10th spatial direction.  We also introduce the notation $\hat e^{i_1}$, $\hat e^{i_2} $, and $ \hat e^{i_3} $ for the frames on the unit radius spaces $AdS_3$, $S^3_2$, and $S^3_3$ respectively.  The 11-dimensional frames may then be written in terms of $\hat e^{i_1}$, $\hat e^{i_2} $, and $ \hat e^{i_3} $ as follows
\begin{align}
e^{i_1} = f_1 \hat e^{i_1} & \hskip 1in &  i_1=0,1,2
\no \\
e^{i_2} = f_2 \hat e^{i_2} &&  i_{2}=3,4,5
\no \\
e^{i_3} = f_3 \hat e^{i_3} &&  i_{3}=6,7,8
\no \\
e^a \hskip 0.57in &&  a=9,\natural
\end{align}
In terms of the frame components, the ansatz for the field strength is
\begin{align}
F =  g_{1 a} e^{012a} + g_{2 a} e^{345a} + g_{3 a} e^{678a}
\end{align}
with the corresponding gauge potential given by
\begin{align}
C = b_1 \hat e^{012} + b_2 \hat e^{345} + b_3 \hat e^{678}
\end{align}
We use the shorthand notation $e^{ij...k} \equiv e^i \wedge e^j \wedge ... \wedge e^k$.

We introduce complex coordinates $w$, $\bar w$ on $\Sigma$, so that the metric takes the form $ds^2_\Sigma = 4 \rho^2 |dw|^2$.  In terms of real coordinates $x$ and $y$ defined by $w = x+i y$, the metric is given by $ds^2_\Sigma = 4 \rho^2 (dx^2 + dy^2)$.  The frames $e^a$ may then be expressed in terms of $x$ and $y$ as $e^9 = 2 \rho \, dx$ and $e^\natural = 2 \rho \, dy$.  We also introduce a complex frame $e^z$ on $\Sigma$ defined by
\begin{align}\label{complexframe}
&e^z = (e^9 + i e^{\natural} )/2 = \rho \, dw& &e^{\bar z} = (e^9 - i e^{\natural} )/2= \rho \, d \bar w&
\end{align}
in terms of which the metric becomes  $ds^2 _{\Sigma} = 4 e^z e^{\bar z}$.
The field strength components can be written with frame or coordinate indices as
\begin{align}
g_{iz} = g_{i9} - i g_{i \natural} = {1\over \rho}g_{iw} ={1\over 2\rho}( g_{ix} - i g_{i y})
\end{align}

In general requiring the existence of non-vanishing $\ep$ which satisfy the BPS equation \eqref{fullgravitino} places restrictions on the supergravity fields (see \cite{Gauntlett:2002fz,Gauntlett:2003wb} for a comprehensive analysis).  In the case we consider here, the supergroup $D(2, 1; c'; 0)\oplus D(2, 1; c'; 0)$ has sixteen independent supersymmetries and so we will require the BPS equation to have sixteen linearly independent solutions.  In this case, the BPS equation will completely constrain the metric factors $f_1$, $f_2$, $f_3$, $\rho$ (up to a sign), and the field strength components $g_{1a}$, $g_{2a}$, $g_{3a}$\footnote{We will be able to completely determine all the physical fields i.e. metric factors and fields strength components in terms of functions $h$ and $H$ (see \eqref{Hdef}) using the BPS constraints only. However, it will turn out that the Bianchi identities and the Einstein equations will add the constraint that $Im(H)\leq0$. It is possible that this constraint could be found from the BPS equations only, by plugging the solutions back into these equations. We did not attempt this.}.

To implement the $D(2, 1; c'; 0)\oplus D(2, 1; c'; 0)$ symmetry, we first decompose the supersymmetry parameter $\ep$ as a tensor product of Killing spinors on the unit radius space $AdS_3$, $S_1^3$ and $S_2^3$ as
\begin{align}
\ep = \sum_{\eta_1, \eta_2, \eta_3}
\chi^{\eta_1, \eta_2, \eta_3} \otimes \zeta_{\eta_1, \eta_2, \eta_3}
\end{align}
$\chi ^{\eta_1, \eta _2, \eta _3}$ are 8-component Killing spinors on the symmetric part of the space, i.e. they are tensor products of the
2-component Killing spinors on $AdS_3$, $S^3_2$, and $S^3_3$. There are eight $\chi ^{\eta_1, \eta _2, \eta _3}$, labeled by $\eta_1=\pm$, $\eta_2=\pm$, and $\eta_3=\pm$. The eight corresponding $\zeta_{\eta_1, \eta_2, \eta_3}$ are 4-component spinors.
From the Majorana condition on $\ep$, the spinors can be shown to obey the following reality conditions
\begin{align}\label{reality}
(\chi^{\eta_1, \eta_2, \eta_3})^*
= (I_2 \otimes \sigma^2 \otimes \sigma^2) \chi^{\eta_1, \eta_2, \eta_3} \qquad \qquad
(\zeta_{\eta_1, \eta_2, \eta_3})^* = (\sigma^3 \otimes \sigma^1) \zeta_{\eta_1, \eta_2, \eta_3}
\end{align}
Due to symmetries of the reduced BPS equations, the eight  $\zeta_{\eta_1, \eta_2, \eta_3}$ can all be related to each other by the following operations.
\begin{align}
&S_0 : \zeta \, \to \, i (I \otimes \sigma ^3) \, \zeta
&\mbox{  which has the effect of  }&
&\eta _i \to - \eta _i &
 \\
 &S_j : \zeta \, \to \, s_j (\sigma ^j  \otimes I) \, \zeta
&\mbox{  which has the effect of  }&
&\eta _i \to - (-)^{\delta _{i,j}} \eta _i  &
\mbox{  and  }
g_{ia} \to - (-)^{\delta _{i.j}}  g_{ia} \no
\end{align}
where $j=1,2,3$, with $s_1=i$, and $ s_2=s_3=1$.
$S_0,\, S_1$, $S_2$, and $S_3$ generate the full symmetry group $\{ I, \, S_0, \, S_1, \, S_0S_1, \, S_2, \, S_0S_2, \, S_3, \, S_0 S_3 \}$ which relates all eight $\zeta_{\eta_1, \eta_2, \eta_3}$. Moreover, the four components of $\zeta _{+++}$ (and any other $\zeta_{\eta_1, \eta_2, \eta_3}$) are reduced by the reality condition (\ref{reality}) to two independent complex components, $\a$ and $\b$
\begin{align}
\zeta _{+++}= \left( \begin{array}{c} \ba \\ -\bb \\ \a \\ \b \end{array} \right )
\end{align}

The BPS equation in the directions of the symmetric spaces reduces to six algebraic conditions on $\a$ and $\b$. Three of these are solved to yield expressions for the metric factors in terms of $\a$ and $\b$
\begin{align}\label{metric}
&f_1=\frac{1}{c_1}(|\a|^2+|\b|^2)&  &f_2=-\frac{1}{c_2}(|\a|^2-|\b|^2)&     &f_3=\frac{i}{c_3}(\a\bb-\ba\b)&
\end{align}
where $c_1$, $c_2$, and $c_3$ are real integration constants. These integration constants are directly related to the $c'$ parameter of the $D(2, 1; c'; 0)$ algebra by the relation $c'=c_2/c_3$ \cite{D'Hoker:2008ix}.
Two more of the algebraic conditions lead to equations relating the fluxes $g_{1z}$, $g_{2z}$ and $g_{3z}$
\begin{align}\label{algebraic}
g_{1z}(\a^2-\b^2)=2(c_1-c_3)+2ig_{3z}\a\b \cr
g_{2z}(\a^2+\b^2)=2(c_2-c_3)+2ig_{3z}\a\b
\end{align}
The last algebraic condition yields the constraint
\begin{align}
c_1+c_2+c_3=0
\end{align}
As a consequence of this constraint, there is only one independent parameter out of the three $c_i$.  The constraint removes one degree of freedom and an overall rescaling of all three parameters can be absorbed into the definition of $\a$ and $\b$.  We will often use the parameter $c=(c_1-c_2)/c_3$ as the remaining parameter.  This parameter is, of course, uniquely related to the $c'$ in $D(2, 1; c'; 0)$ by the formula $c=-2c'-1$.

The BPS equation in the directions along $\Sigma$ is reduced to the following differential equations on $\a$ and $\b$.
\begin{align}
\label{diffab}
&D_z \a   =  + { i \over 2} \hat \o_z \a +
{1 \over 12} g_{1z} \b - {1 \over 12} g_{2z} \b - { i \over 12} g_{3z} \a
\no \\
&D_z \b  =  + { i \over 2} \hat \o_z \b +
{1 \over 12} g_{1z} \a + {1 \over 12} g_{2z} \a + { i \over 12} g_{3z} \b
\no \\
&D_z \ba   =  - { i \over 2} \hat \o_z \ba +
{1 \over 4} g_{1z} \bb - {1 \over 4} g_{2z} \bb + { i \over 4} g_{3z} \ba
\no \\
&D_z \bb   =  - { i \over 2} \hat \o_z \bb +
{1 \over 4} g_{1z} \ba + {1 \over 4} g_{2z} \ba - { i \over 4} g_{3z} \bb
\end{align}
where $\hat \o_z$ is the spin connection on $\Sigma$ and we have introduced the notation
\begin{align}
D_z={\partial_w\over\rho}\hspace*{20pt}D_{\bar z}={\partial_{\bar w}\over\rho}
\end{align}
as implied by (\ref{complexframe}). For vanishing torsion, the frame is related to the spin connection by $e^a + \omega^a {}_b \wedge e^b = 0$, which can be used to compute $\hat \o_z$ (the $z$-component of $\omega^8 {}_9$) in terms of $\rho$.
\begin{align}
\label{omegadef}
\hat \o_z=i {\partial_w \rho\over\rho^{2}}dw \hspace*{20pt}
\hat \o_{\bar z}=-i {\partial_{\bar w} \rho\over\rho^{2}}d\bar w
\end{align}

Finally, it was shown in \cite{D'Hoker:2008wc} that the system of equations \eqref{diffab} after using the algebraic constraint \eqref{algebraic} admit a first integral for any value of $c$.  We will put off giving this first integral until the next section where we will rederive it.  In \cite{D'Hoker:2008wc}, the full solutions of the system of equations defined by \eqref{algebraic} and \eqref{diffab} were found only for the special values of $c = 0,3,-3$.  The values $c=0,3$ correspond to solutions asymptotic to $AdS_7 \times S^4$, while the value $c = -3$ corresponds to solutions asymptotic to $AdS_4 \times S^7$.  However, we note that the reductions reviewed in this section and the form of the first integral are all valid for general $c$.  In the remainder of this paper, we solve the system of equations for general values of $c$.

\section{Solution for General Values of $c$}\label{sectionreduction}
\setcounter{equation}{0}

In this section, we derive the solutions to the system of equations defined by \eqref{algebraic} and \eqref{diffab} for general values of $c$.
This will require a different method than the one used in \cite{D'Hoker:2008wc} for the special values of $c$.

\subsection{Reduction to two equations and the first integral}\label{sectionreduction1}

Our first task is to eliminate the $g_i$'s and reduce the system of four BPS equations
to a system of two equations, one of which is algebraic in $\rho$ and inhomogenous in both $\alpha$ and $\beta$, while the other is differential in $\rho$ and homogeneous in both $\alpha$ and $\beta$.
We also recover the aforementioned first integral found in \cite{D'Hoker:2008wc}.   We emphasize that it is not necessary to use the first integral to eliminate variables in order to obtain the two equation system.

We begin by introducing $\as$ and $\bs$ by rescaling the variables $\alpha$ and $\beta$ in the following way
\begin{align}
\label{scaledvars}
\begin{split}
\alpha={1\over \rho^{{1\over 4}}}\left({\bar \kappa^3\over\kappa}\right)^{{1\over 8}}\as  \qquad
\bar\alpha={1\over \rho^{{1\over 4}}}\left({ \kappa^3\over\bar\kappa}\right)^{{1\over 8}}\abs \cr
\beta={1\over \rho^{{1\over 4}}}\left({\bar \kappa^3\over\kappa}\right)^{{1\over 8}}\bs \qquad
\bar\beta={1\over \rho^{{1\over 4}}}\left({\kappa^3\over\bar\kappa}\right)^{{1\over 8}}\bbs
\end{split}
\end{align}
where $\kappa$ is an arbitrary holomorphic function. We have chosen the above factor of $\rho$ so that $\rho$ drops out of the first two equations in \eqref{omegadiff}.  The rescaling by $\kappa$ is just a notational convenience\footnote{
Note that in  \cite{D'Hoker:2008wc}, $\kappa$ is used to denote the first integral. We also adopt this notation later on, but at this point in the calculation, $\kappa$ is just an arbitrary holomorphic function. }.

Next we rewrite the differential equations \eqref{diffab} as logarithmic derivatives for the combinations $\o_{\a}=\as\bar\as^3$, $\o_{\b}=\bs\bar\bs^3$, $\bar\o_{\a}=\bar\as\as^3$ and $\bar\o_{\b}=\bar\bs\bs^3$.  Although, we could express $\alpha$, $\beta$ in terms of these new variables, we do not do so for now.  These definitions are made by demanding that both $\pw\rho$ and $g_{3z}$ drop out of the equations for $\oab$, and $\obb$.
\begin{align}
\begin{split}\label{omegadiff}
\pw\ln\oab&={1\over 4} \rho(g_{1z}-g_{2z})\left({\bs\over\as}+{\bbs\over\abs} \right)\cr
\pw\ln\obb&={1\over 4} \rho(g_{1z}+g_{2z})\left({\as\over\bs}+{\abs\over\bbs} \right)\cr
\pw\ln\oa&=2\pw\ln\rho-\pw\ln\k+{1\over 4}(g_{1z}-g_{2z})\rho\left( 3{\bbs\over\abs}+{1\over 3}{\bs\over\as} \right)+i {2\over 3}\rho g_{3z}\cr
\pw\ln\ob&=2\pw\ln\rho-\pw\ln\k+{1\over 4}(g_{1z}+g_{2z})\rho\left( 3{\abs\over\bbs}+{1\over 3}{\as\over\bs} \right)-i {2\over 3}\rho g_{3z}
\end{split}
\end{align}

We now proceed to exclude the $g_{iz}$'s which will leave us with three equations.  First, we rewrite the algebraic constraints (\ref{algebraic}) to have one equation without $g_{3z}$ and another to be homogenous in the $g_{iz}$ (i.e. without the constant term)
\begin{align}
\begin{split}
\label{algebraic2}
&(g_{1z}-g_{2z})\alpha^2-(g_{1z}+g_{2z})\beta^2=2(c_1-c_2) \cr
&g_{1z}(\a^2-\b^2)(c_2-c_3)-g_{2z}(\a^2+\b^2)(c_1-c_3)+2ig_{3z}\a\b(c_1-c_2)=0
\end{split}
\end{align}
Next we obtain an inhomogenous differential equation without $\pw\rho$ and a homogenous differential equation.  To do so, we use the first two equations of (\ref{omegadiff}) to express the combinations $(g_{1z}+g_{2z})$ and $(g_{1z}-g_{2z})$ in terms of $\pw\ln\oab$ and $\pw\ln\obb$. To obtain the inhomogenous differential equation, we insert these expressions into the first equation of \eqref{algebraic2}.  To obtain the homogenous differential equation, we add the last two equations of  \eqref{omegadiff} which eliminates $g_{3z}$ and then eliminate the expressions for $(g_{1z}+g_{2z})$ and $(g_{1z}-g_{2z})$ (using the first two equations of (\ref{omegadiff}) as before).  The resulting two differential equations are
\begin{align}
\begin{split}
\label{systemo}
\pw\oab-\pw\obb=&{\kappa\over 2}(c_1-c_2)\left({\rho\over|\kappa|}\right)^{3\over 2}(\abs\bs+\as\bbs) \cr
{\pw\oa\over\oa}+{\pw\ob\over\ob}=&4\pw\ln {\rho\over|\kappa|}
+{1\over 3}{\pw\oab\over\oab}{ 9\bbs\as+\bs\abs\over\abs\bs+\as\bbs}+{1\over 3}{\pw\obb\over\obb}{ 9\abs\bs+\as\bbs \over\abs\bs+\as\bbs}
\end{split}
\end{align}

There then remain two more independent equations. One of them must give the expression for $g_{3z}$.  $g_{3z}$ is obtained using the difference of the $\oa$ and $\ob$ equations, i.e. the last two equations of \eqref{omegadiff}, but this equation is not displayed.  We get the final equation by inserting this expression for $g_{3z}$ into the second equation of \eqref{algebraic2} and once again insert the expressions for $(g_{1z}+g_{2z})$ and $(g_{1z}-g_{2z})$.  The final result is a total derivative
\begin{align}\label{scaledconstraint}
\p_w\Big((\abs\as^3-\bs\bs^3)-{c_1-c_2\over c_3}\as\bs(\abs\bs-\as\bbs)\Big)=0
\end{align}
The above equation implies that the expression in parenthesis is a first integral.
%Note that we did not need to assume any form, in order to obtain this first integral.\JE{I don't like this sentence.}
Using the fact $\pw\bar\kappa=0$, the above equation implies that
\begin{align}
\p_w\Bigg\{\rho\Big((\ba\a^3-\bb\b^3)-{c_1-c_2\over c_3}\a\b(\ba\b-\a\bb)\Big)\Big\}=0
\end{align}
In other words, the expression in parenthesis above is an anti-holomorphic function. Now, since the function $\bar\kappa$ used in the rescaling \eqref{scaledvars} was an arbitrary anti-holomorphic function, we can choose
\begin{align}\label{kappaconstraint}
\bar\kappa=\rho((\ba\a^3-\bb\b^3)-{c_1-c_2\over c_3}\a\b(\ba\b-\a\bb))
\end{align}
After this identification, we see that $\bar\kappa$ is a first integral.  This is also the expression for the first integral which was derived in \cite{D'Hoker:2008wc}. Note that, as was observed there, $\kappa$ is a $(1,0)$ form. It is interesting to note that while $\a$ and $\b$ are $(-1/4,1/4)$ forms, the quantities $\as$ and $\bs$, as well as their complex conjugates, are $(0,0)$ forms.

Before we proceed, we would like to discuss how the case of $c_1 = c_2$ (i.e. $c=0$) fits into the story.  The careful reader may have observed that in this case the inhomogeneous differential equation, \eqref{systemo}, reduces to \eqref{scaledconstraint}!  This is a consequence of the fact that the algebraic constraints \eqref{algebraic2} become degenerate.  However, this is not actually a problem.  Since this is simply an intermediate step in the calculation, we can proceed for general $c$ and take the limit $c\rightarrow 0$ once the full solution is found and recover the $c=0$ solution found in \cite{D'Hoker:2008wc}.  Indeed, it will turn out that all of the physical variables of the problem (metric factors, field strength components, spinor components) are expressed in terms of a harmonic function, $h$, and another function, $G$, which satisfies a simple partial differential equation.  All of these expressions and the partial differential equation will be valid for any value of $c$.

We note that it is also certainly possible to follow a derivation which is valid for any $c$ at every step.  A different inhomogeneous equation can be found by going back to an earlier form of the algebraic constraints (\ref{algebraic}) which are not degenerate. We do not pursue this course here because the inhomogenous equation obtained this way is more complicated for general $c$ and we already have a derivation for $c=0$ valid at every step in \cite{D'Hoker:2008wc}.

We now continue working with our equations. We can rewrite the constraint \eqref{kappaconstraint} in terms of the rescaled variables $\as$ and $\bs$, or alternatively in terms of $\oa$ and $\ob$
\begin{align}
\begin{split}
\label{kappaconstraint2}
1&=(\abs\as^3-\bar\bs\bs^3)-{c_1-c_2\over c_3}\as\bs(\abs\bs-\as\bbs) \cr
1&=\oab\Big(1+{c_1-c_2\over c_3}\left|{\ob\over \oa}\right|^{1\over 2}\Big)-\obb\Big(1+{c_1-c_2\over c_3}\left|{\oa\over \ob}\right|^{1\over 2}\Big)
\end{split}
\end{align}
Since we have two constraints, (\ref{kappaconstraint}) and its complex conjugate, we should be able to rewrite our system (\ref{systemo}) in terms of only two variables. To this end, we introduce a new variable $\l^4=\ob/\oa$ and its complex conjugate, in terms of which the second equation of \eqref{kappaconstraint2} can be written as
\begin{align}
\oab=\frac{1}{(1+c\l\bar\l)-\bar\l^4(1+c(\l\bar\l)^{-1})}\label{omegalambda}
\end{align}
where as mentioned before $c=(c_1-c_2)/ c_3$.  This will allow us to eliminate $\ob$, $\oa$, $\oab$ and $\obb$ in terms of $\lambda$ and $\bar \lambda$.  We can now easily express equations (\ref{systemo}) in terms of $\ob$, $\oa$, $\oab$, $\obb$ using
\begin{align}\label{alphabetaomega}
\bar\alpha_s\beta_s&=\left({\oa^3\over\oab} \right)^{1\over 8}\left({\obb^3\over\ob} \right)^{1\over 8}=\oa^{1\over 4}\oab^{1\over 4}\bar\l^{3\over 2}\l^{-{1\over 2}}\no\\
\as\bs&=\left({\oab^3\over\oa}\right)^{1\over 8}\left({\ob^3\over\ob}\right)^{1\over 8}={\oab^{3\over 4}\over\oa^{1\over 4}}{\l^{3\over 2}\over\bar\l^{1\over 2}}
\end{align}
The equations \eqref{systemo} in terms of $\lambda$ and $\oa$ and their conjugates are
\begin{align}\label{systemolambda}
\pw\oab-\pw(\bar\l^4\oab)&={\kappa\over 2}(c_1-c_2)\left({\rho\over|\kappa|}\right)^{3\over 2}\oa^{1\over 4}\oab^{1\over 4}{\bar\l^2+\l^2\over(\l\bar\l)^{1\over 2}}\\
{3\over 4}\pw\ln(\l^2\oa)&=\pw\ln\left({\rho\over |\kappa|}\right)^{3\over 2}+{1\over 8}\pw\ln\oab{\bar\l^2+9\l^2 \over \bar\l^2+\l^2}+{1\over 8}\pw\ln(\bar\l^4\oab){9\bar\l^2+\l^2 \over \bar\l^2+\l^2}\no
\end{align}
This system of equations can now be expressed in terms of only $\l$, $\bar\l$, and $\rho$ by plugging in \eqref{omegalambda} and its complex conjugate into \eqref{systemolambda}. However, we do not write down these expressions explicitly because they are unwieldy and unilluminating.
Instead we now explain in general terms, the strategy we use to perform an integration, executing the analytic but messy algebraic computations in Mathematica.

\subsection{Integrating out $\rho$}
The system of equations (\ref{systemolambda}) can be put into the following general form
\begin{align}
\label{systemolambda2}
C(\l,\bar\l)\pw\l+D(\l,\bar\l)\pw\bar\l&={c_1-c_2\over 2}\kappa \left({\rho\over|\kappa|}\right)^{3\over 2}\no\\
A(\l,\bar\l)\pw\l+B(\l,\bar\l)\pw\bar\l&=\pw\ln\left({\rho\over|\kappa|}\right)^{3\over 2}
\end{align}
In  this subsection,
we show that we can rewrite \eqref{systemolambda2} as a completely equivalent system, in which $\rho$ is dressed by a multiplicative factor, $M(\l,\bar\l)$.  In this equivalent system, the differential forms on the left hand sides in terms of $\l$ and $\bar\l$ are equal and the system can be integrated.  The choice of the functional form of $M(\l,\bar\l)$, in terms of $\l$ and $\bar\l$, will be explained below.
The resulting system is
\begin{align}\label{systemform}
M(C\pw\l+D\pw\bar\l)&={c_1-c_2\over 2}\kappa \left({\rho\over|\kappa|}\right)^{3\over 2}M\no\\
(A+\p_{\lambda}\ln(M ))\pw\l+(B+ \p_{\bar\lambda}\ln(M   ))\pw\bar\l&=\pw\ln\Big(M\left({\rho\over|\kappa|}\right)^{3\over 2}\Big)
\end{align}
Note that the system \eqref{systemform} is completely equivalent to \eqref{systemolambda2}. All we have done is multiplied the first equation by
the function $M(\l,\bar\l)$, while in the second line we added to both sides the quantity $\p_w \ln M(\l,\bar \l)$.  The only case in which these two systems are inequivalent is the case of singular $M$, i.e. when it is $0$ or $\infty$, but since we will construct $M$ explicitly, it will be clear that this does not occur.  Since the left hand sides are equal for such an $M$, the right hand sides must be equal as well and the following easily integrable equation is obtained.
\begin{align}\label{systemmiracle}
{c_1-c_2\over 2}\kappa\left({\rho\over|\kappa|}\right)^{3\over 2}M=\pw\ln\Big(M\left({\rho\over|\kappa|}\right)^{3\over 2}\Big)
%M(C\pw\l+D\pw\bar\l)=(A+\p_{\lambda}\ln(M ))\pw\l+(B+ \p_{\bar\lambda}\ln(M   ))\pw\bar\l
\end{align}

We now show that we can construct $M(\l,\bar\l)$ such that the left hand sides are equal and moreover that the coefficients of $\p_w\l$ and $\p_w\bar\l$ are equal.   Note that for a completely general system of the form (\ref{systemolambda2}) i.e. for arbitrary functions $A$, $B$, $C$, and $D$,
there is no guarantee that an $M$ can be found which equates the left hand sides of (\ref{systemform}).  However, $A$, $B$, $C$, and $D$ are not arbitrary here, but specific functions of $\l$, $\bar\l$, determined from (\ref{systemolambda}).
Equating the coefficients of $\p_w\l$ and $\p_w\bar\l$ in (\ref{systemform}) yields the two equations
\begin{align}\label{diffeqm}
{1\over M}{\p M\over\p \l}+A=C M\qquad\qquad  {1\over M}{\p M\over \p\bar\l}+B=D M
\end{align}
%\begin{align}\label{systemform}
%M(\l,\bar\l)(C(\l,\bar\l)\pw\l+D(\l,\bar\l)\pw\bar\l)=M(\l,\bar\l)\rho^{3\over 2}\\\no\\
%(A(\l,\bar\l)+\p_{\lambda}\ln(M(\l,\bar\l) )\pw\l+(B(\l,\bar\l)+ \p_{\bar\lambda}\ln(M(\l,\bar\l)   )\pw\bar\l=\no\\
%M(\l,\bar\l)(C(\l,\bar\l)\pw\l+D(\l,\bar\l)\pw\bar\l)=\pw\ln(M(\l,\bar\l)\rho^{3\over 2})\no
%\end{align}
Fortunately, we do not even have to solve these differential equations to find $M$ %(assuming it exists)
since it must satisfy both differential equations and is therefore overdetermined.\footnote{This is the main obstruction in constructing $M$ for arbitrary $A$, $B$, $C$, $D$.}
We next construct $M$ algebraically as follows. Note that, upon obtaining $M$ from the algebraic expressions, one must still check that (\ref{diffeqm}) are satisfied since the system is overdetermined.  To derive the algebraic expressions, we linearize (\ref{diffeqm}) by dividing by $M$ and rewriting the equation in terms of the new variable $N=1/M$
\begin{align}\label{Nderiv}
-{\p N\over\p \l}+AN=C \qquad\qquad  -{\p N\over \p\bar\l}+BN=D
\end{align}
We then differentiate the resulting equations such that both contain the second order mixed derivative of $N$, and use (\ref{Nderiv}) to exclude the first order derivatives of $N$. The second order derivative cancels between the two equations and the following expression is obtained
\begin{align}
{1\over M}={A D-B C+{\p C\over\p\bar\l}-{\p D \over\p\l}  \over  {\p A\over\p\bar\l}-{\p B \over\p\l}  }
\end{align}
Applying this procedure to (\ref{systemolambda}) with (\ref{omegalambda}) inserted yields the following expression for $M$.
%Note that (\ref{1overM}) indeed satisfies (\ref{diffeqm}). %(Note that if the multiplier can be multiplied by any constant without changing its effect.)
\begin{align}\label{1overM}
{1\over M}=-{c\over 2}{(\l^2-\bar\l^2)(\l^2\bar\l^2-1)\over (\l\bar\l)^{1\over 2}}\left({\l\bar\l\over (\l+c\l^2\bar\l-c\bar\l^3-\l\bar\l^4)(\bar\l+c\bar\l^2\l-c\l^3-\bar\l\l^4)}\right)^{3\over 4}
\end{align}
One can check that (\ref{1overM}) indeed satisfies both equations in (\ref{diffeqm}) and thus provides the advertised multiplier.

Although this expression is quite ugly, in subsection \ref{subsec:compdisc}, it will be shown to be something interesting and possibly geometric.
We note that $M$ is imaginary. Also note that $M$ is guaranteed to exist and not to be 0 or $\infty$. The only way $M$ can be 0 is if   $(\l+c\l^2\bar\l-c\bar\l^3-\l\bar\l^4)=0$ which restricts $\l$, $\bar\l$ to be constant.  For $M$ to be $\infty$, either $\l$ has to be real or $|\l|=1$ which are both trivial cases (e.g. this implies $\rho=0$).
Next, we discuss what the equations have become. The equation \eqref{systemmiracle} for $\rho$ can now be easily solved by rewriting it as
\begin{align}
\pw\left( \left( M\left({\rho\over |\kappa|}\right)^{3\over 2}\right) ^{-1}\right)= - \kappa {c_1-c_2\over 2}
\end{align}
Since $\kappa$ is holomorphic, we can integrate both sides.  To do so explicitly, we introduce the imaginary harmonic function $h$ by
\begin{align}
\pw h=-\kappa \label{hkappa}
\end{align}
Integrating, we then have
\begin{align}
\label{hdef}
\left(M\left({\rho\over |\kappa|}\right)^{3\over 2}\right)^{-1} = h {c_1-c_2\over 2}
\end{align}
Finally we are left with one very ugly equation to solve.  To obtain it, we take the second line of \eqref{systemolambda2} and use \eqref{hdef} to eliminate $\rho$
\begin{align}\label{oneeqn}
\pw\ln h&=A_2\pw\l+B_2\pw\bar\l\\
A_2&=-{1\over\bar\l-\l}+{1\over \bar\l+\l}+{1\over \l(\l\bar\l-1)}-{1\over\l(\l\bar\l+1)}+{2c\bar\l(\l^2+\bar\l^2)\over \l(-\l-c\l^2\bar\l+c\bar\l^3+\l\bar\l^4)}\no\\
B_2&={1\over\bar\l-\l}+{1\over \bar\l+\l}+{\l\over \l\bar\l-1}+{\l\over\l\bar\l+1}+{2(c\l^2-3c\bar\l^2-4\l\bar\l^3)\over -\l-c\l^2\bar\l+c\bar\l^3+\l\bar\l^4}+{4\bar\l  \over \l^2+\bar\l^2}\no
\end{align}
Since $\kappa$ is an arbitrary holomorphic function, $h$ is an arbitrary harmonic function that we choose. The choice of $h$ determines $\kappa$ and the above equation is solved for a given $h$.
This final equation may then be viewed as a differential equation for $\l$ and $\bar \l$.

\subsection{Linear equation}

Our next goal is to integrate \eqref{oneeqn}.
Even though \eqref{oneeqn} depends on $c$ in $\lambda$ variables, we show below that it can be mapped to a quasi linear equation which is independent of $c$.
In fact, this differential equation is the same as the one obtained for the special values of $c = -3,0,3$ in \cite{D'Hoker:2008wc}.

Life would be easy if the right hand side of \eqref{oneeqn} would be a total derivative, but this is not the case.
The next best thing that can be tried
is to multiply \eqref{oneeqn} by some multiplier, $m$, such that the right hand side becomes a total derivative of some function, which we call $G(\lambda,\bar \lambda)$ and furthermore that $m$ is a linear function of $G$ and $\bar G$ (the latter is why the equation is only quasi-linear).
In \cite{D'Hoker:2008wc}, an equation of this form was obtained by different methods and for special values of $c$.
\begin{align}\label{Geqn}
\pw G={1\over 2}(G+\bar G)\pw\ln h
\end{align}
The above equation was shown in \cite{D'Hoker:2008wc} to be valid for all 3 values of $c\in\{0,3,-3\}$(, although each case was separately derived). This leads one to guess that the equation is the same for all $c$. In order for this to be the case, comparing \eqref{oneeqn} with \eqref{Geqn}, we see that $m=(1/ 2)(G+\bar G)$\footnote{More generally, just from demanding linearity, one could try the ansatz $m = a G + b \bar G$ with $a$ and $b$ arbitrary complex constants.  However, this is more complicated and not necessary for the problem considered here.}.

We describe our procedure to integrate the equation in general terms but the algebra is performed in Mathematica.
Before we start, we note that, as can be seen below in \eqref{linearizingintconst}, the numerical coefficient in front of $G+\bar G$ drops out of the integrability conditions and one can look for $m$ in the more general form $m = \mbox{constant}*(G+\bar G)$.  However, the procedure picks out a unique coefficient when the final expression for $G(\l,\bar\l)$ is plugged into (\ref{Geqn}) and compared with (\ref{oneeqn}).  For our case, one finds that the coefficient is indeed $1/2$.
After multiplying \eqref{oneeqn} by $(1/ 2)(G+\bar G)$ and equating the result with \eqref{Geqn}, one obtains
\begin{align}
\pw G=&\p_{\l}G\pw \l+\p_{\bar\l}G\pw\bar\l={1\over 2}A_2(G+\bar G)\pw \l+{1\over 2}B_2(G+\bar G)\pw\bar\l
\end{align}
Equating the coefficients of $\pw \l$ and $\pw \bar \l$ we obtain the conditions
\begin{align}\label{solveforG}
&\p_{\l}G={1\over 2}(G+\bar G)A_2  \qquad \qquad \qquad \p_{\bar\l}G={1\over 2}(G+\bar G)B_2
\end{align}

The system of equations, \eqref{solveforG}, is an overdetermined system.  Similar to the case of looking for $M$, once we find $G$, using the integrability conditions for the system \eqref{solveforG}, we must check that $2 \p_w G/(G+\bar G)$ is actually equal to the right hand side of (\ref{oneeqn}). To derive the integrability conditions, we take the mixed derivative of \eqref{solveforG} and equate the second order derivatives.  We also note that the first order derivatives are simply related.
\begin{align}
\label{linearizingintconst}
\p_{\l}\p_{\bar\l}G&={\p\over\p\bar\l}(A_2{1\over 2}(G+\bar G))={\p\over\p\l}(B_2{1\over 2}(G+\bar G))\\
\p_{\l}G&={A_2\over B_2}\p_{\bar\l}G
\end{align}
Note also that both equations can be complex conjugated providing additional equations.
Solving these four equations together, we obtain
\begin{align}\label{Gderiv}
\p_{\l}\bar G=(G+\bar G){\bar B_2(\p_{\l}B_2-\p_{\bar\l}A_2)\over A_2\bar A_2 - B_2\bar B_2 }=(G+\bar G)C_2\no\\
\p_{\l}G=(G+\bar G){A_2(\p_{\bar\l}\bar B_2-\p_{\l}\bar A_2)\over A_2\bar A_2 - B_2\bar B_2 }=(G+\bar G)D_2
\end{align}
where we have defined new variables $C_2$ and $D_2$. Adding the above equations, we obtain a simple equation and solution for $G+\bar G$.
\begin{align}
{\p \over \p \l}(G+\bar G)&=(G+\bar G)(C_2+D_2)\no\\
G+\bar G&=Exp\Big(\int\big({\bar B_2(\p_{\l}B_2-\p_{\bar\l}A_2)\over A_2\bar A_2 - B_2\bar B_2 }+{A_2(\p_{\bar\l}\bar B_2-\p_{\l}\bar A_2)\over A_2\bar A_2 - B_2\bar B_2 }\big)d\l)\Big)f(\bar\l)\no\\
G+\bar G&={i(\l^4-\bar\l^4)(1-\l^2\bar\l^2)\over (\bar\l(1-\l^4)+c\l(\bar\l^2-\l^2))(\l(1-\bar\l^4)+c\bar\l(\l^2-\bar\l^2))  }
\end{align}
where $f(\bar\l)$ is an arbitrary function which has appeared as a result of the integration. Fortunately, we are able to remove this degree of freedom
because the expression for $G+\bar G$ has to be real. The result of the integration actually has some branch cuts and the correct choice using this function is just to have the $i$ in the numerator as shown. We still have the freedom to multiply by an arbitrary real constant but that is just a symmetry of equation (\ref{Geqn}). To find $G-\bar G$ we must subtract the equations of (\ref{Gderiv}) and integrate again.
\begin{align}\label{gminusgbar}
{\p\over \p\l}(G-\bar G)&=(D_2-C_2)(G+\bar G)\no\\
G-\bar G&={-i(\l^2-\bar\l^2)^2(1+2c\l\bar\l+\l^2\bar\l^2)  \over  (\bar\l(1-\l^4)+c\l(\bar\l^2-\l^2))(\l(1-\bar\l^4)+c\bar\l(\l^2-\bar\l^2))}
\end{align}
Note that we had the freedom of an additive function of $\bar\l$ in the second equation of (\ref{gminusgbar}) coming from the integration,
but as before this freedom is reduced to that of adding a constant because we know that this expression has to be pure imaginary.
This constant is fixed to be $i/c$ and is already included in the above expression. We fixed this constant by demanding that the expression has a finite limit as $c\rightarrow 0$ since we know the solution to be perfectly well behaved at this point from the results of \cite{D'Hoker:2008wc}. We also note that equation (\ref{Geqn}) allows us to add an arbitrary imaginary constant to $G$. $G$ is given by the following.
\begin{align}\label{G}
G=i{(\bar\l+\l)\bar\l(\l-\bar\l)   \over \l+c\l^2\bar\l-c\bar\l^3-\l\bar\l^4 }
\end{align}
With this identification we can check that (\ref{G}) inserted into (\ref{Geqn}) exactly reproduces (\ref{oneeqn}). Since this equation is independent of $c$ and identical to what was obtained in \cite{D'Hoker:2008wc}, the integration method used there (in section 8) to integrate (\ref{Geqn}) applies to our work as well. However, the range of $G$ and the expressions of the physical fields in terms of $G$ do depend on $c$ as is discussed in section \ref{subsec:ranges}.

\subsection{Comparison with previous results and discussion}\label{subsec:compdisc}

We would now like to compare our results to those of \cite{D'Hoker:2008wc} for the special values of $c$ that are solved there. To this end, we will express $G$ in terms of the original variables $\a$, $\b$, $\rho$.
\begin{align}\label{Galphabeta}
G={i \a\b(\ba\b-\a\bb)\over \ba\a^3-\bb\b^3-c\a\b(\b\ba-\a\bb)}=i \a\b(\ba\b-\a\bb)\rho/\bar\kappa=i \as\bs(\abs\bs-\as\bbs)
\end{align}
where to obtain the second equality we notice that the denominator is $\bar\kappa/\rho$.
We must also express the $G$ obtained for $c=0$ and $c=-3$ in terms of $\a$, $\b$, $\rho$. In the case of $c=0$ we get a straight-forward agreement.
For the case of $c=-3$, we find that $G$ appearing in \cite{D'Hoker:2008wc} has the following form as a function of $\alpha$, $\beta$.
\begin{align}\label{Gcminus3}
G_{c=-3,\cite{D'Hoker:2008wc}}=i{\ba\a^3-\bb\b^3+\a\b(\a\bb-\ba\b)\over \ba\a^3-\bb\b^3+3\a\b(\b\ba-\a\bb)}=i{(\ba\a^3-\bb\b^3+\a\b(\a\bb-\ba\b))\over\bar\kappa/\rho}
\end{align}
This expression looks quite different from (\ref{Galphabeta}) but it is actually equivalent up to  a real multiplicative constant and an imaginary additive constant.  As we have already noted, adding an imaginary constant or multiplying by a real constant is a symmetry of G since solutions differing in this way are not distinguished by (\ref{Geqn}).  (We also note here that due to branch cuts we can only claim to be able to express $G$ in terms of $\a$, $\b$ up to a multiplicative 4th root of 1. This is also true for the $G$ expression (\ref{Galphabeta}). However, once we obtain the metric factor formulas, we will see that the above choice of branch cut yields $AdS_7\times S^4$ and $AdS_4\times S^7$.) To show the equivalence of (\ref{Gcminus3}) and (\ref{Galphabeta}) we rewrite (\ref{Gcminus3}) by adding and subtracting $i$.
\begin{align}\label{Gcminus32}
G_{c=-3,\cite{D'Hoker:2008wc}}&=i{(\ba\a^3-\bb\b^3+\a\b(\a\bb-\ba\b))-(\ba\a^3-\bb\b^3+3\a\b(\b\ba-\a\bb))\over  \ba\a^3-\bb\b^3+3\a\b(\b\ba-\a\bb)}+i\no\\&=-4i\a\b(\b\ba-\a\bb)\rho/\bar\kappa+i
=-4G+i
\end{align}
We would now like to emphasize that amazingly, not only is $G$ of the form (\ref{Galphabeta}), which is rather simple even for general $c$, but also the dependence on $c$ can be hidden inside of $\rho$ or inside of the scaled variables $\as$, $\bs$.
%We now pause and take a look at our results. It seems at first glance that we have traveled very far from our original variables and that we have %generalized a great deal as compared to the cases of special values of $c$. However, as we are about to see, there are remarkable and simple %relations to the original variables and the special case problem. It requires an unpleasant amount of algebra but it is quite possible to express %our results in terms of the original variables $\a$, $\b$.
%Not only is the result remarkably simple but it is also unchanged from the result for the case of special $c$ values except that on must remember %to use the full $\kappa$ constraint with general $c$.
It is also interesting to look at the form of $M$ in terms of $\a$ and $\b$.
\begin{align}
{1\over M}&={c(\a\ba+\b\bb)(\a\ba-\b\bb)(\a\bb-\ba\b)\over
2 \Big((\ba\a^3-\bb\b^3-c\a\b(\ba\b-\a\bb))(\a\ba^3-\b\bb^3-c\ba\bb(\a\bb-\ba\b))\Big)^{3\over 4}}\label{Malphabeta}\\
&={-i c c_1 c_2 c_3  f_1 f_2 f_3 \over 2 (|\kappa|/\rho)^{3\over 2}}\label{Mmetric}
\end{align}
where to get the last equality we have used the expressions for the metric factors (\ref{metric}).
  Again, it is remarkable how simple this expression is and again the dependence on $c$ can be hidden inside of $\rho$.
Now if we remember the relationship between $h$, $M$, and $\rho$, (\ref{hdef}), and we insert (\ref{Mmetric}) we uncover another constraint.
\begin{align}\label{hmetrics}
h&={2\over c_1-c_2}{1\over M}\left({|\kappa|\over\rho}\right)^{3\over 2}\no\\
&={i c c_1 c_2 c_3 \over c_1-c_2}f_1 f_2 f_3=i c_1 c_2 f_1 f_2 f_3
\end{align}
In \cite{D'Hoker:2008wc}, it was also noted (for the special $c$ values) that the product of the metric factors is a harmonic function. However, note that one does not need to find the metric factors as functions of $G$ to prove this fact.
Also note that we can rediscover the definition of $\kappa$ and get an independent check of (\ref{hkappa}). If we differentiate $h$ using the second equality of (\ref{hmetrics}) and use only the original equations (\ref{algebraic}-\ref{diffab}), we recover $\kappa$.
\begin{align}
\p_w h=\p_w({i c c_1 c_2 c_3 \over c_1-c_2}f_1 f_2 f_3)=- \rho (\a\ba^3-\b\bb^3-c\ba\bb(\a\bb-\ba\b))=-\kappa
\end{align}
%The above equation defines $\rho$.
%\begin{align}
%\rho=\p_w h/(\a\ba^3-\b\bb^3-c\ba\bb(\a\bb-\ba\b))
%\end{align}
%We emphasize here that the free parameter of our solution is $h$

\section{Summary of Solution and Remarks}\label{sectionsummary}
\setcounter{equation}{0}

In this section we give a summary of the general solution.  We then give a discussion on the allowed range of the function $G$.  Finally, we show how $AdS_4 \times S^7$ and $AdS_7 \times S^4$ fit into the framework of this paper and give examples of one parameter deformations as a function of $c$ for each case.

\subsection{Summary of the general solution}

The explicit expressions for the metric factors are derived in Appendix \ref{sec:metric} and the corresponding formulas for the fluxes are derived in Appendix \ref{sec:fluxes}.  To help simplify the final expressions, we introduce the real harmonic function $\hat h$ related to the imaginary harmonic function $h$ by $h = i \hat h$.  The solutions are then specified by the choice of a Riemann surface $\Sigma$ and the triple $\{c,G,\hat h\}$, where
$\hat h$ is a real harmonic function on $\Sigma$, and $G$ is a complex function which satisfies the differential equation
\begin{align}
\p_w G = \frac{1}{2} (G + \bar G) \p_w \ln \hat h
\end{align}
while $c$ is a real constant, which can take any real value.  $c$ is related to the $c_i$ appearing in the expressions for the metric factors \eqref{metric} by $c = (c_1 - c_2)/c_3$.  The $c_i$ satisfy a constraint $c_1 + c_2 + c_3 = 0$, while an overall rescaling of the $c_i$ can be absorbed into $\hat h$ so that one may always fix one of the $c_i$ to be one.  Therefore specifying the value of $c$ is sufficient to determine the $c_i$.
The differential equation for $G$ can be solved in terms of an integral after picking $\hat h$ as a local coordinate \cite{D'Hoker:2008wc}.

In addition to satisfying a differential equation, there is also a restriction on the range of allowed values for $G$, which is derived in Appendix \ref{sec:metric}.
The constraint is derived and expressed most easily in terms of $H = G/(1-i c G)$
\footnote{Alternatively, one may derive the first equation of \eqref{finalrangeconst} directly from the expression for $G$ in terms of $\lambda$ and $\bar \lambda$ given in \eqref{G}.  However, the restriction to imaginary values of $H$ follows from the equations of motion.}
\begin{align}
\label{finalrangeconst}
&4|H|^4 + (H-\bar H)^2 \leq 0 \cr
&\textrm{Im}(H)\leq0
\end{align}
The range of allowed values for $G$ is further discussed in section \ref{subsec:ranges}.

The metric is given by
\begin{align}
ds^2 = f_1^2 ds^2_{AdS_3} + f_2^2 ds^2_{S^3} + f_1^2 ds^2_{S^3} + 4 \rho^2 dw d\bar w,
\end{align}
where the metric factors are given by
\begin{align}\label{rhofinalH}
\rho^6&={|\p_w \hat h|^6\over  c_3^4 \hat h^4 |1+icH|^{6}} (1-|H|^2)\Big(- 4|H|^4 - (H-\bar H)^2\Big)
\end{align}
\begin{align}
\label{f1finalH}
f_1^6&={c_3^2 \over c_1^6}\hat h^2(1-|H|^2)\frac{\big(2|H|^2+ i(H-\bar H)\big)^3}{\Big(4|H|^4+(H-\bar H)^2\Big)^2}
\end{align}
\begin{align}
\label{f2finalH}
f_2^6&=- {c_3^2 \over c_2^6}\hat h^2(1-|H|^2)\frac{\big(2|H|^2- i(H-\bar H)\big)^3}{\Big(4|H|^4+(H-\bar H)^2\Big)^2}
\end{align}
\begin{align}
\label{f3finalH}
f_3^6&={\hat h^2\over c_3^4} \left( \frac{-(H-\bar H)^2-4|H|^4}{(1-|H|^2)^2} \right)
\end{align}
A particularly simple expression is given by the following product of metric factors
\begin{align}
\label{simpexp}
f_1 f_2 f_3 = \frac{\hat h}{c_1 c_2}
\end{align}

The field strength is given by
\begin{align}
F = (d b_1) \wedge \hat e^{012} + (d b_2) \wedge \hat e^{345} + (d b_3) \wedge \hat e^{678}
\end{align}
where $\hat e^{012}$ is the unit volume form on $AdS_3$ and $\hat e^{345}$ and $\hat e^{678}$ are unit volume forms on $S^3$, while $d b_1$, $d b_2$ and $d b_3$ are given by
\begin{align}
\label{g1Hfinal}
\p_w b_1=& 2 \nu_1 \frac{c_3}{c_1^3} i \hat h
\Bigg[\bar H(-i+c H)(H-3 \bar H+4H \bar H^2) \p_w H - H\bar H(H-3 \bar H+4H \bar H^2) \p_w H \no\\
& -i \, c\Big((H-\bar H)^2-H\bar H^3+3H^2\bar H^2 \Big) \p_w H +H(-i+c H)(H+\bar H)(1-i H) \p_w\bar H \Bigg]
\no\\&
\Bigg((-i+c H)(H+\bar H)\Big( 2 H\bar H-i(H-\bar H) \Big) ^2  \Bigg)^{-1}
\end{align}
\begin{align}
\label{g2Hfinal}
\p_w b_2=&-2 \nu_2 \frac{c_3}{c_2^3} i \hat h
\Bigg[\bar H(-i+c H)(H-3 \bar H+4H \bar H^2) \p_w H + H\bar H(H-3 \bar H+4H \bar H^2) \p_w H
\no\\&
+ i \, c \Big((H-\bar H)^2-H\bar H^3+3H^2\bar H^2 \Big) \p_w H + H (-i+c H)(H+\bar H)(1+i H) \p_w\bar H \Bigg]
\no\\&
\Bigg((-i+c H)(H+\bar H)\Big( 2 H\bar H+ i(H-\bar H) \Big) ^2  \Bigg)^{-1}
\end{align}
\begin{align}
\label{g3Hfinal}
\p_w b_3 =&
\frac{\nu_3}{c_3^2} \hat h
\Bigg[ \Bigg( 8 i H^2 \bar H^3 + 3 c H \bar H^3 - i \bar H^3 + 3 c H^2 \bar H^2 - 15 i H \bar H^2 - 2 c \bar H^2 + 2 i H^2 \bar H - c H \bar H
\no\\ & \qquad \qquad
+ 5 i \bar H + c H^2 -3 i H  \Bigg) \p_w H + (-i+c H)(1+H^2)(H+\bar H) \p_w \bar H + \Bigg]
\no\\&
\Bigg((-i+c H)(H+\bar H)(1 - H \bar H)^2  \Bigg)^{-1}
\end{align}
where $\nu_i = \pm 1$.  The equations of motion determine the fluxes up to an overall sign.  In order to determine the signs, we must require the fluxes and metric factors to satisfy the BPS equations.  This can be done for the special cases $c = \{-3,0,3\}$ where we take the geometry to be either $AdS_4 \times S^7$ ($c=-3$) or $AdS_7 \times S^4$ ($c=0,3$).  Using the explicit formula for $G$ and $\hat h$ given in section \ref{sec:explicit} to compute the fluxes and comparing to the expressions given in section 3 of \cite{D'Hoker:2008wc}, we find $\nu_1 = 1$, $\nu_2 = -1$ and $\nu_3 = -1$.  Additionally, as reviewed in section \ref{sectionsetup}, the BPS equations admit a symmetry where one may flip the signs of any two fluxes.  Conversely, when we flip the sign of a single flux, the BPS equations do not admit any solutions.  Thus we conclude that the configurations with $\{\nu_1,\nu_2,\nu_3\} = \{+,+,+\}, \{+,-,-\}, \{-,-,+\}, \{-,+,-\}$ all preserve supersymmetry, while the configurations with $\{\nu_1,\nu_2,\nu_3\} = \{+,+,-\}, \{+,-,+\}, \{-,+,+\}, \{-,-,-\}$ all spontaneously break supersymmetry.  This way of breaking supersymmetry is similar to the "skew-whiffed" solutions of \cite{Duff:1983nu,Duff:1986hr}

There are two ways of realizing $AdS_7 \times S^4$ in the local solution.  The first is realized by the following choice of $\{c,G,\hat h\}$
\begin{align}
\label{AdS7exp}
c& =0&
G &= i \frac{\sinh(w-\bar w)}{\sinh(2\bar w)}& \cr
\hat h &= - i \frac{L^3 c_3^2}{4} \cosh(2w) + c.c.&
H &= G&
\end{align}
where $0 \leq \Re(w) \leq \infty$ and $0 \leq \Im(w) \leq \pi/2$ and $L$ is the radius of the $S^4$.  The other $AdS_7 \times S^4$ solution is given by taking
\begin{align}
c& =3&
G &= - \frac{i}{4} \frac{\sinh(w-\bar w)}{\sinh(2\bar w)} - \frac{i}{4} & \cr
\hat h &= -  i (2 L c_3^2) \cosh(2w) + c.c.&
H &= - i \frac{\sinh(w-\bar w)+ \sinh(2 \bar w)}{\sinh(2 \bar w) - 3 \sinh(w-\bar w)} &
\end{align}
The $AdS_4 \times S^7$ solution is given by
\begin{align}
\label{AdS4exp}
c& =-3&
G &= - \frac{i}{4} \frac{\cosh(w+\bar w)}{\cosh(2\bar w)} + \frac{i}{4}& \cr
\hat h &= i \frac{L^3 c_3^2}{4} \sinh(2w) + c.c.&
H &= i \frac{\cosh(2 \bar w) - \cosh(w + \bar w)}{\cosh(2 \bar w) + 3 \cosh(w + \bar w)}&
\end{align}
where $-\infty \leq \Re(w) \leq \infty$ and $0 \leq \Im(w) \leq \pi/2$ and $L$ is the radius of the $S^7$.

An arbitrary choice of $\{c,G,\hat h\}$ will generically lead to solutions which contain singularities.  It is therefore interesting to ask what types of choices lead to either regular geometries or geometries with physically allowed singularities, such as those caused by brane sources.
In the case of $AdS_4 \times S^7$ or $AdS_7 \times S^4$, it turns out that $\Sigma$ has a boundary, but that boundary does not correspond to a boundary in the 11-dimensional space-time.  Rather one of the $S^3$'s always vanishes on the boundary of $\Sigma$, leading to a smooth capping off of the geometry.  Requiring this structure, we find that \eqref{simpexp} implies that $\hat h = 0$ on the boundary of $\Sigma$.  Furthermore, if we also require $f_1$ to remain finite on the boundary of $\Sigma$, we must then have $4|H|^4+(H-\bar H)^2 = 0$ so that \eqref{f1finalH} can remain finite.  In other words, on the boundary of $\Sigma$, $H$ must take values in the boundary of its range.
%Finally, in order for the cap to be smooth, we need only one of the sphere factors to vanish, while the other remains finite.  From \eqref{f2finalH} and \eqref{f3finalH}, this requires $H$ to be either $0$ or $-i$ on the boundary of $\Sigma$.
While we have not shown that these boundary conditions are sufficient to guarantee smoothness of the geometry, we see that they are necessary.  More generally, one may also consider various types of singularities.  We leave both of these issues for future work, but give examples of novel smooth solutions in section \ref{sec:explicit}.

\subsection{Range of $G$ and $H$} \label{subsec:ranges}

\begin{figure}
\begin{center}
\includegraphics[width=0.3 \linewidth]{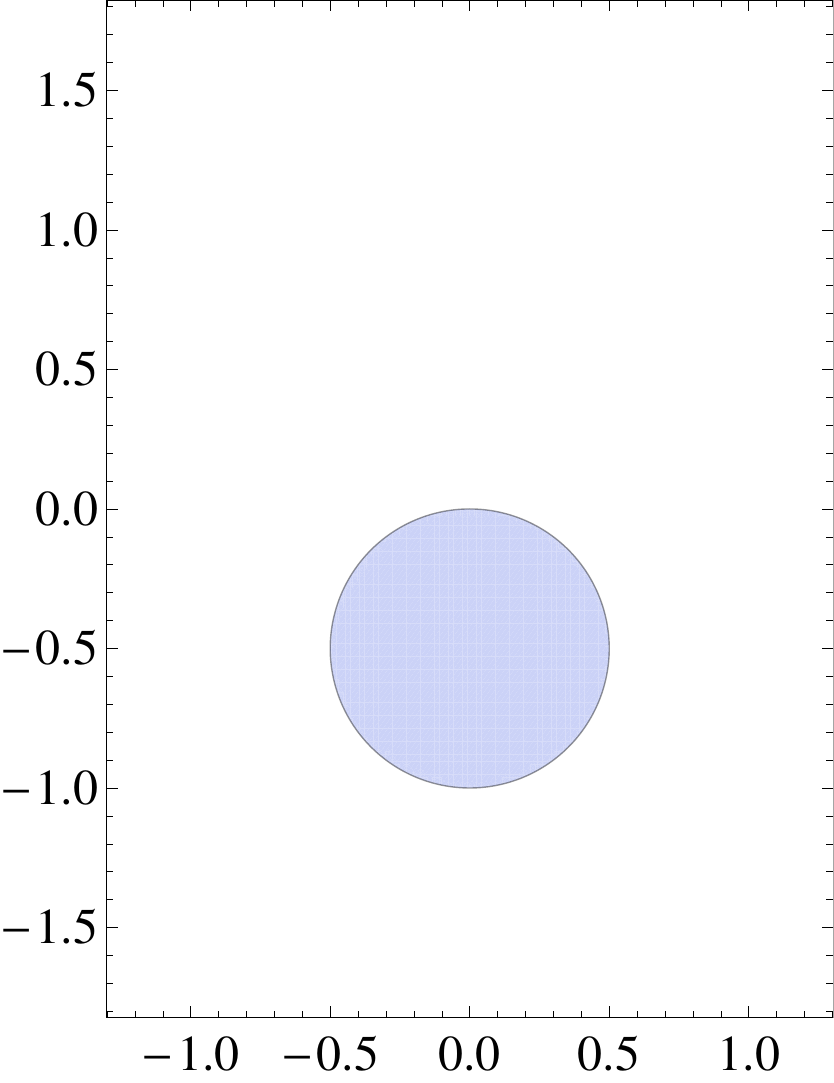}
\caption{Range of $H$ for any $c$.}
\label{Hrange}
\end{center}
\end{figure}

It is interesting to determine the range of the functions $G$ and $H$.  Note that the range of $H$, depicted in Fig. \ref{Hrange}, is fixed and independent of $c$ (modulo a small technical point when $c=0$, which will be explained shortly).  As discussed above \eqref{f1start}, the range of $H$ is determined by the condition $|H-\bar H|\geq 2|H|^2$.
%The weaker condition $|H|\leq 1$ is implied by this previous condition.
In general, this allows for two branches of $H$, one with $\textrm{Im}(H)\geq0$ and one with $\textrm{Im}(H)\leq0$.  However, as discussed in Appendix \ref{sec:fluxes}, the Bianchi identity for $g_{3z}$ selects the branch with $\textrm{Im}(H)\leq0$, except in the special case $c=0$.

The reason $c=0$ is special, is because the solutions corresponding to the mirror image range for which $i(H-\bar H)\leq 0$ are just an automorphism of the old range for which $i(H-\bar H)\geq 0$.  This can be seen as follows. An equivalent solution is one for which the solution $G$ to its differential equation (\ref{Geqn}) is the same up to the symmetries discussed below (\ref{Gcminus3}) and the expressions for the physical fields are unchanged. Consider the mapping $H\rightarrow -H$ which implies $G\rightarrow -G$ and $c\rightarrow -c$. The change of sign in $G$ is undetected by the differential equation which is also independent of $c$. However, for general $c$ the expressions for the metric factors change in two ways: we must pick the other sign in front of $i(H-\bar H)$ and we must exchange $c_1  \leftrightarrow c_2$ (where the latter is the only way to map $c \rightarrow -c$). For general $c$, this new combination differs from the original and is not a solution.
%In the expression for the physical fields, picking the other sign in front of $i(H-\bar H)$ undoes the negation of $H$ but then it would be odd if switching $c_1$ and $c_2$ was a new solution in addition to the original one.
However, for the special case $c=0$, we have $c_1=c_2$ and so the solution is simply mapped into itself.  Thus the two branches are equivalent.

We note that the restriction to $\textrm{Im}(H)\leq0$ is consistent with the results of \cite{D'Hoker:2008wc}.  In the case of $c=-3$, corresponding to $AdS_4 \times S^7$ asymptotics, the range of $G$ consisted of only a single branch.  For the case of $c=0$, corresponding to $AdS_7 \times S^4$, there were two branches for the range of $G$, one with $\textrm{Im}(G) > 0$ and one with $\textrm{Im}(G)<0$, which intersect at $G=0$.  One might wonder whether there are solutions where $G$ interpolates between the two allowed regions, which intersect at $G = 0$.  However, the explicit solutions constructed in \cite{D'Hoker:2008qm} indicate that this is not possible.

\begin{figure}
\begin{center}
    \includegraphics[width=0.3 \linewidth]{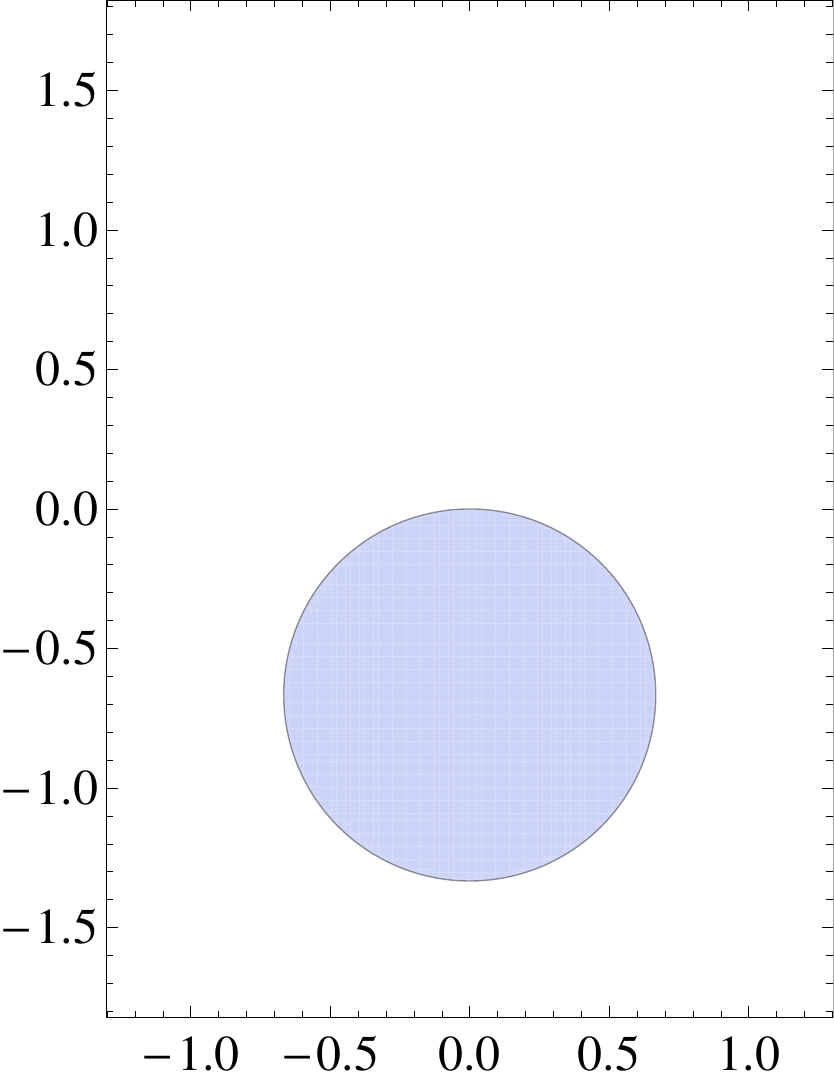}
    \includegraphics[width=0.3 \linewidth]{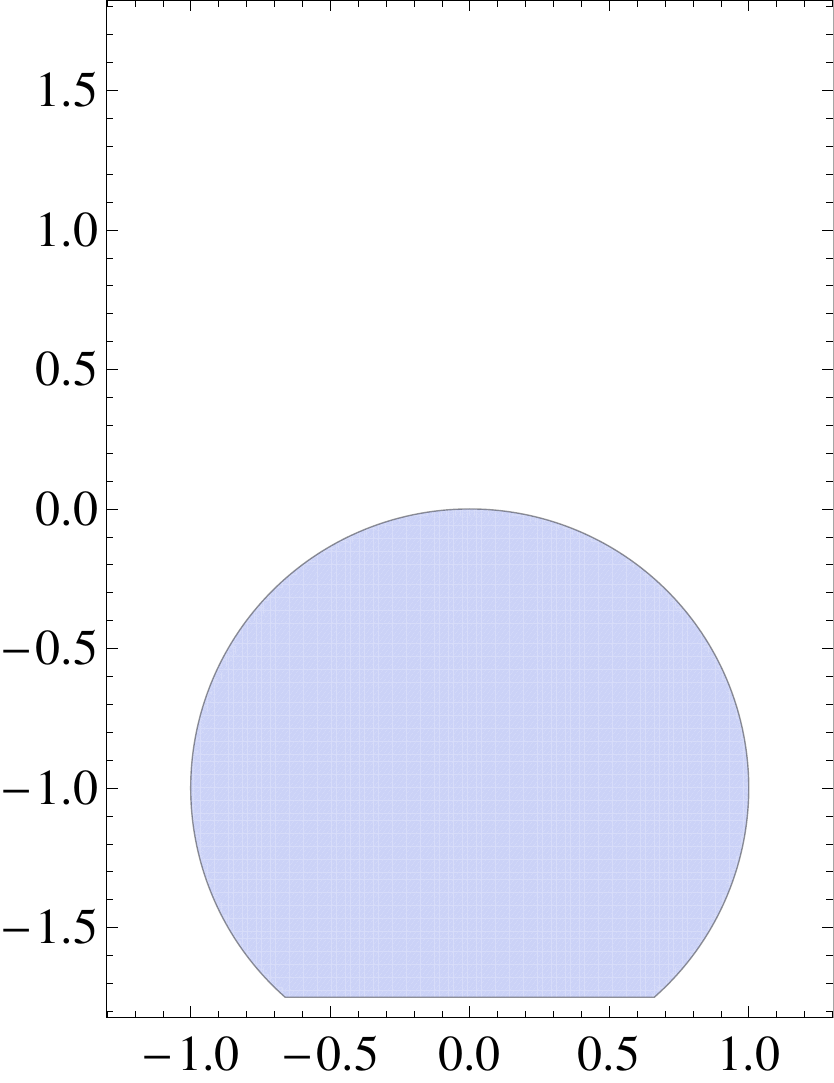}
    \includegraphics[width=0.3 \linewidth]{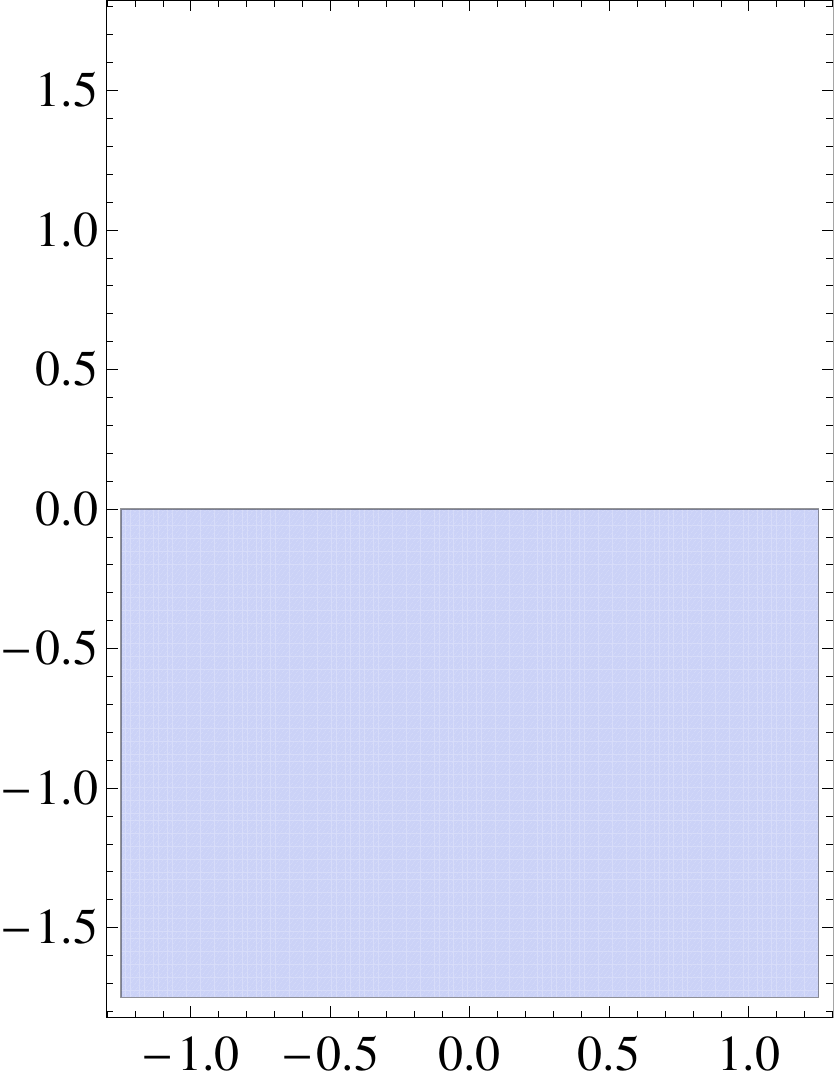}
    \caption{Range of $G$ for $c=-.25$ (left), $c=-.5$ (middle) and $c=-1$ (right)}
 \label{Grangeminuspoin25tominusp1}
 \end{center}
\end{figure}

The quantity $H$ is convenient to work with precisely because its range is independent of $c$. However, the differential equation obeyed by $H$ is not linear and depends on $c$. Since we will eventually have to impose boundary conditions for the differential equation of $G$, it is interesting to consider the range of $G$ as well (which is most easily obtained from mapping the range of $H$).
For $c=0$, we have $G=H$ and the range of $G$ is of course simply the range of $H$ given in Fig. \ref{Hrange}. As $c$ is taken to be more negative, the radius of the allowed range circle increases until it decompactifies at $c=-1$ so that any $G$ with a negative imaginary part is allowed.  This sequence is illustrated in Fig. \ref{Grangeminuspoin25tominusp1}.  Note that $c=-1$ implies that $c_2=0$ which means that one of the spheres decompactifies.

\begin{figure}
\begin{center}
     \includegraphics[width=0.3 \linewidth]{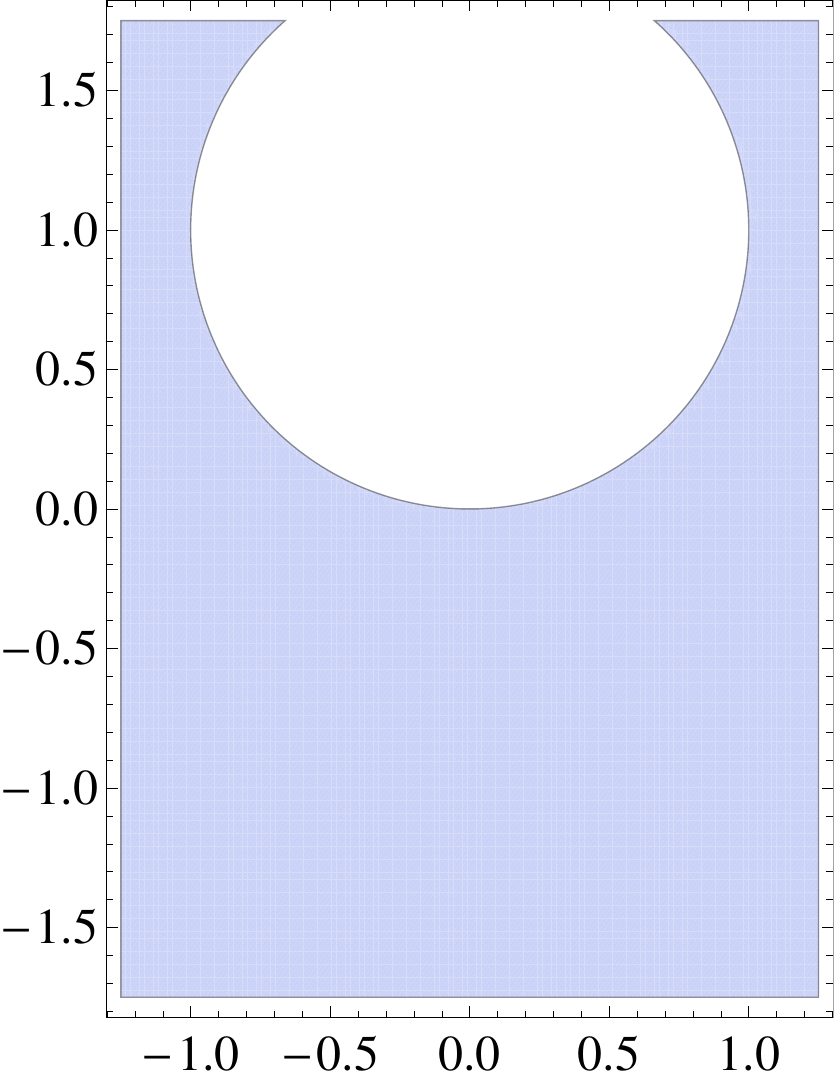}
     \includegraphics[width=0.3 \linewidth]{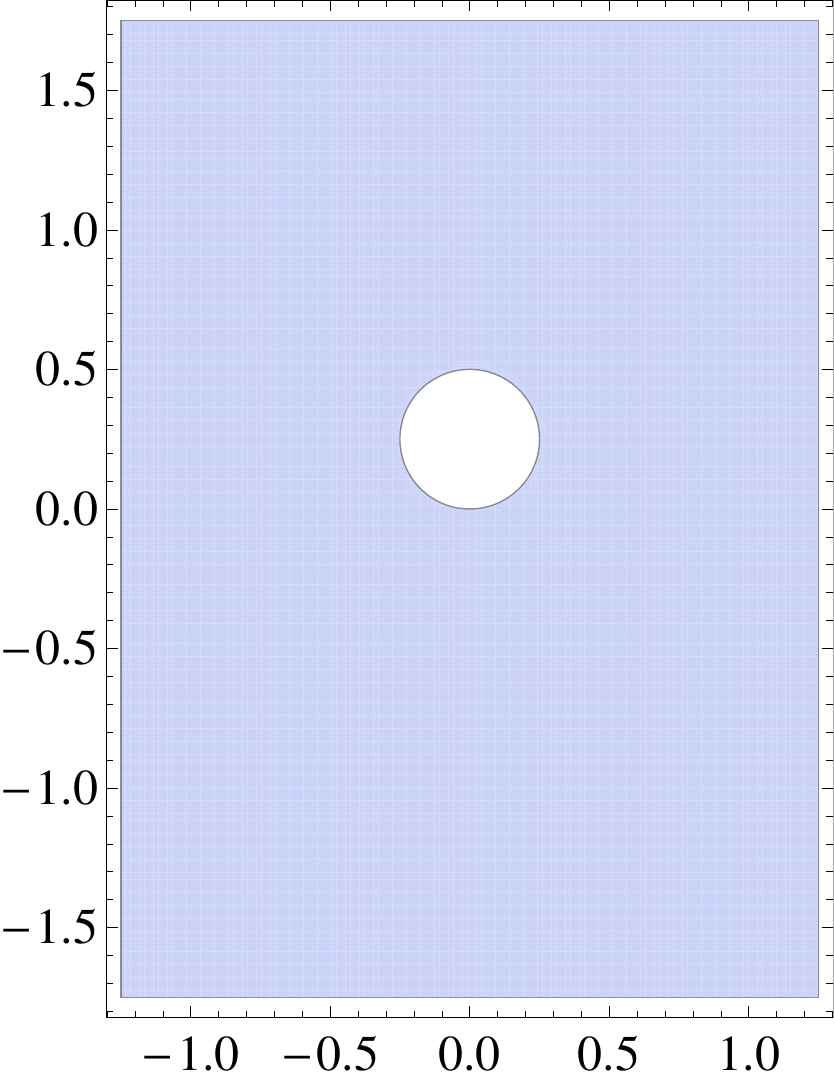}
     \includegraphics[width=0.3 \linewidth]{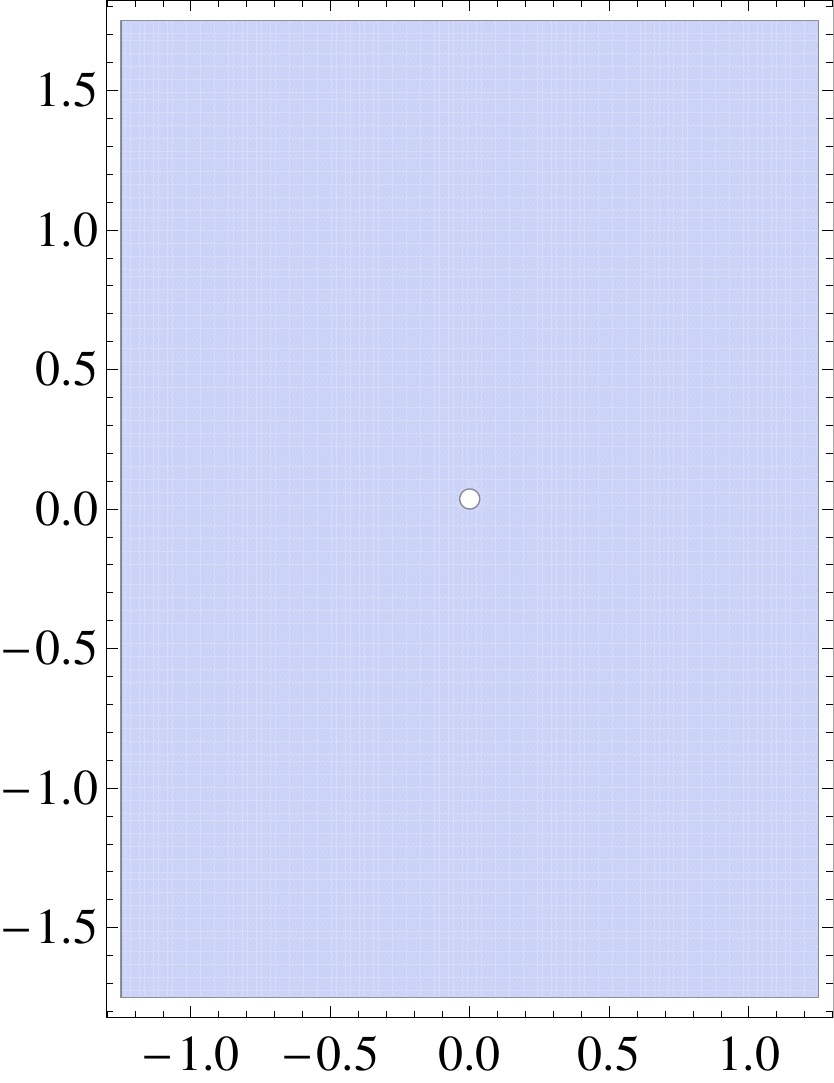}
 \caption{Range of $G$ for $c=-1.5$ (left), $c=-3$ (middle) and $c=-15$ (right).}
 \label{Grangeminus1p5tom15}
 \end{center}
\end{figure}

Continuing to decrease $c$ from $c=-1$, the range further increases such that all $G$ are allowed except a disallowed circle whose radius decreases as $c$ becomes more negative.  This is shown in Fig. \ref{Grangeminus1p5tom15}. Note that the middle figure in Fig. \ref{Grangeminus1p5tom15} is the range for $c=-3$ which includes all solutions asymptotic to $AdS_4\times S^7$.  Our range agrees with the one depicted in figure 2 of section 8 of \cite{D'Hoker:2008wc} if we remember that our $G$ is mapped into that of \cite{D'Hoker:2008wc} by $G_{c=-3,\cite{D'Hoker:2008wc}}=4G+i$ as discussed in \eqref{Gcminus32}.  As $c$ takes values in the range of $\{0,-3\}$ the family of solutions interpolates, as a function of $c$ between solutions asymptotic to $AdS_7\times S^4$ and those asymptotic to $AdS_4\times S^7$.  As $c\rightarrow -\infty$ the range increases as seen in Fig. \ref{Grangeminus1p5tom15} (right) until it includes the entire plane. The case $c=-\infty$ corresponds to $c_3=0$ and is therefore the case of the other sphere decompactifying.

\begin{figure}
\begin{center}
     \includegraphics[width=0.3 \linewidth]{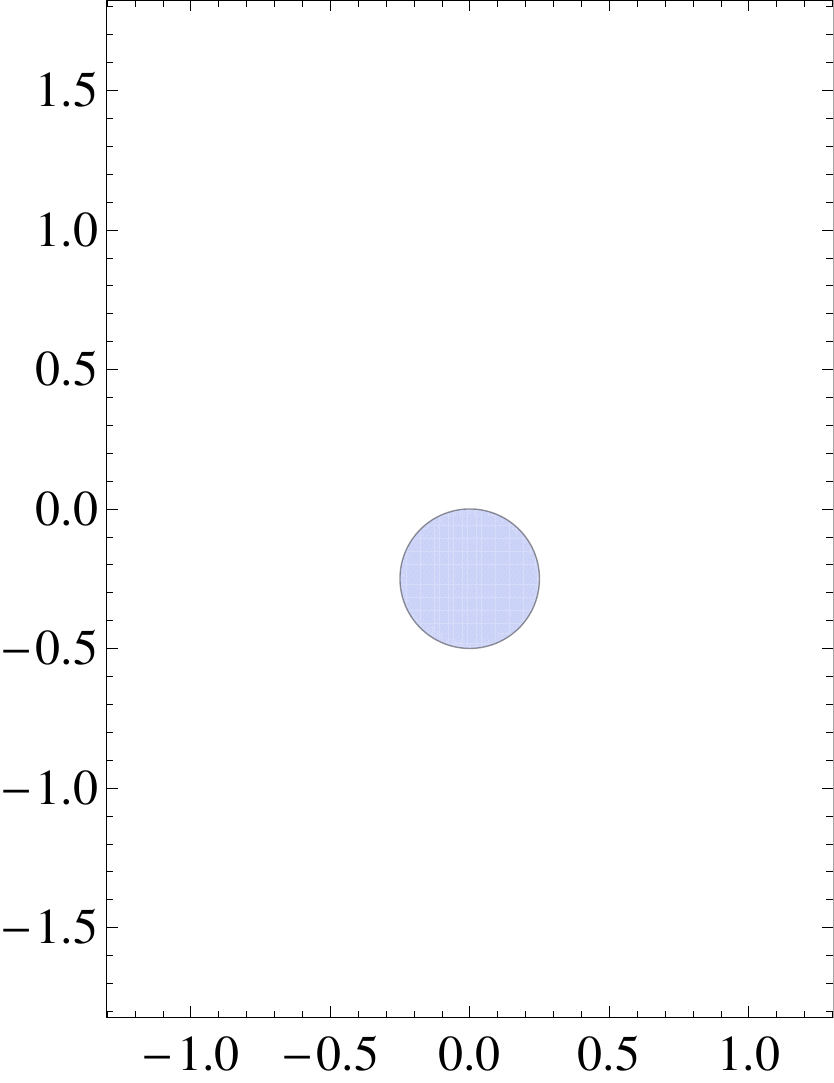}
     \includegraphics[width=0.3 \linewidth]{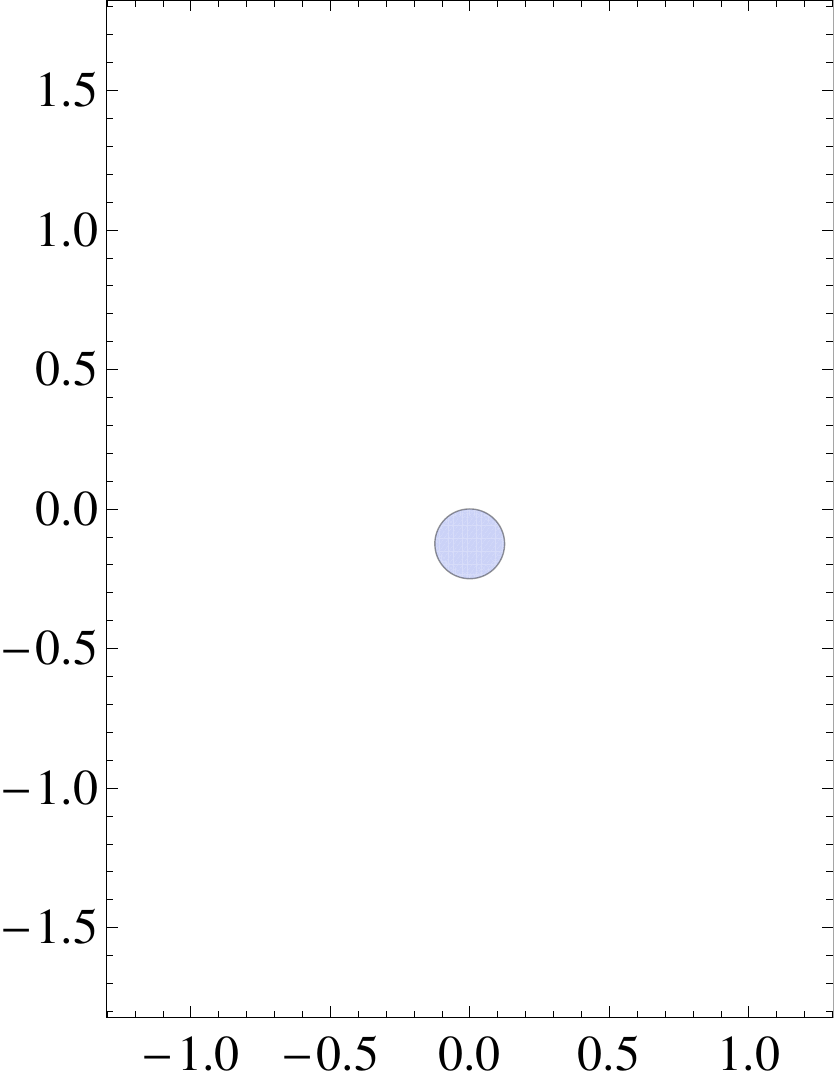}
     \includegraphics[width=0.3 \linewidth]{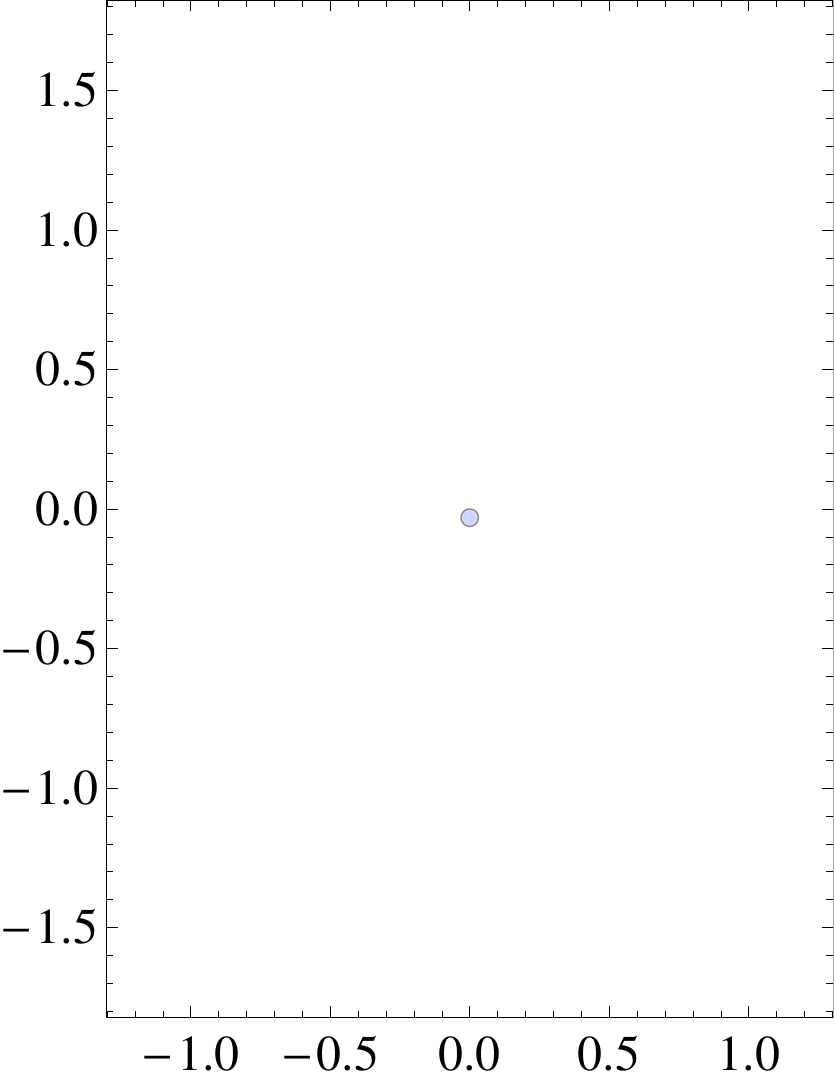}
 \caption{Range of $G$ for $c=1$ (left), $c=3$ (middle) and $c=15$ (right).}
 \label{Grange1to15}
 \end{center}
\end{figure}

We now consider what happens as $c$ increases above 0.  As $c$ grows for $c>0$, the range of $G$ decreases as can be seen in Fig. \ref{Grange1to15}. Note that the $c=3$ case, show in Fig. \ref{Grange1to15} (middle), contains solutions which are asymptotic to $AdS_7\times S^4$ but differs from the $c=0$ case in that the roles of the two 3-spheres are interchanged.  Setting $c=1$ implies $c_1=0$, so this case corresponds to the decompactification of the $AdS_3$ region and is shown in Fig. \ref{Grange1to15} (left).

We summarize the special values of $c$ as follows
\begin{center}
\begin{tabular}{|c|c|c|}
  \hline
%$c = \pm \infty$& $G \in$ plane &$S^3$ decompactification \\
$c = 3$& $G \in $ disc &$AdS_7 \times S^4$ \\
$c = 1$& $G \in $ disc &$AdS_3$ decompactification \\
$c = 0$&  $G \in $ disc &$AdS_7 \times S^4$ \\
$c = -1$ & $G \in$ lower half plane & $S^3$ decompactifiaction \\
$c = -3$ & $G \in $ outside of disc & $AdS_4 \times S^7$ \\ \hline
\end{tabular}
\end{center}
We note that in order to interpolate as a function of $c$ between the two solution classes which asymptote to $AdS_7 \times S^4$ corresponding to $c=3$ or $c=0$, one must pass through solutions with $c=1$ for which the $AdS_3$ part of the metric is decompactified.  Similarly, to interpolate between solutions which are asymptotic to $AdS_4\times S^7$ with $c=-3$ and solutions which are asymptotic to $AdS_7\times S^4$ with $c=0$, one must pass through the solutions with $c=-1$ so that one of the spheres must decompactify.
%To interpolate between the $AdS_4\times S^7$ solutions with $c=-3$ and the $AdS_7\times S^4$ solutions with $c=3$, the other sphere must decompactify at $c=\infty$.  We note that this interpolation requires passing through the point $c=\infty$ and it is likely that one should study this region as a function of $c^{-1}$ or some other well-defined parameter.  There is no obvious relationship between the range of $G$ and the full geometry of the solutions, but we expect that knowing this range will be necessary for finding the global solutions which we intend to work on in the future.

\section{One parameter deformation of $AdS_7 \times S^4$ and $AdS_4 \times S^7$}\label{sec:explicit}
\setcounter{equation}{0}

Here we give one parameter deformations of both $AdS_4 \times S^7$ and $AdS_7 \times S^4$.  We start with $AdS_7 \times S^4$, which is given by the choice of parameters in \eqref{AdS7exp}.
Writing $w=x+iy$, the domain of $\Sigma$ is given by $0 \leq x < \infty$ and $0 \leq y \leq \pi/2$, with the boundary of $AdS_7$ located at $x = \infty$.  One can check that $H$ satisfies the constraint \eqref{finalrangeconst}.\footnote{We note that the $G$ here is the negative of the $G$ given in \cite{D'Hoker:2008wc}.  We remind the reader that in the special case $c=0$, there are two equivalent branches of solutions, while only one branch extends to general values of $c$.}

We now construct a one-parameter deformation of $AdS_7 \times S^4$ as follows.  Keeping $\hat h$ fixed, we first use the real scale and imaginary shift symmetry of the differential equation to write a more general $G$ with two real parameters $a$ and $b$.
\begin{align}
G &= i a \frac{\sinh(w-\bar w)}{\sinh(2\bar w)} + i b
\end{align}
Next we construct $H$ for an arbitrary value of $c$ using $H = G/(1-i c G)$ with the requirement that $H$ satisfies the same boundary conditions as the $c=0$ case so that $H = 0$ when $y = 0$ or $y = \pi/2$ and $H=-i$ when $x=0$.  This leads to the restriction $a = 1/(1+c)$ and $b=0$ with $c$ left as a parameter. Note that for these choices, $H$ satisfies the constraint \eqref{finalrangeconst}.  We leave a fuller discussion of boundary conditions to future work but below we see that the preceding simple choice of boundary conditions on $H$ gives a regular geometry. The corresponding metric factors as a function of $c$ are given by
\begin{align}
f_1^6 &= L^6 \frac{16 \cosh^2(x) [1-c+(1+c)\cosh(2x)][1+c \cos(4y)+(1+c)\cosh(2x)]}{(1+c)(1-c)^6}& \cr
(2 \rho)^6 &= L^6 \frac{16 [1-c+(1+c)\cosh(2x)][1+c \cos(4y)+(1+c)\cosh(2x)]}{(1+c)^4 \cosh^4(x)}& \cr
f_2^6 &= L^6 \frac{128 \cosh^2(x) [1+c \cos(4y)+(1+c)\cosh(2x)] \sinh^6(x) }{(1+c)^4[1-c+(1+c)\cosh(2x)]^2} & \cr
f_3^6 &= L^6 \frac{2 \cosh^2(x) [1-c+(1+c)\cosh(2x)] \sin^6(2y)}{(1+c)[1+c\cos(4y)+(1+c)\cosh(2x)]^2}&
\end{align}
One may readily check that the bulk geometry is regular when $c$ is in the range $-1 < c < 1$.  In particular, the $S^3$'s cap off smoothly at the boundary of $\Sigma$.  When $c$ is in the range $-\infty < c < -1$, the geometry will contain a singularity, which can be seen by noting that $f_1$ will vanish at some value of $x$ and $y$.  The geometry is also smooth for $c$ in the range $1 < c < \infty$.

We now move onto the $AdS_4 \times S^7$ case, which is given by the choice of parameters given in \eqref{AdS4exp}.  Writing $w=x+iy$, the domain of $\Sigma$ is given by $-\infty \leq x < \infty$ and $0 \leq y \leq \pi/2$, with the boundary of $AdS_4$ located at $x = \pm \infty$.  In these coordinates the boundary of $AdS_4$ is split into two halves which are glued together along the $AdS_3$ boundary.  One can check that $H$ satisfies the constraint \eqref{finalrangeconst}.

We can construct a one-parameter deformation of $AdS_4 \times S^7$ using the same technique as before.  Keeping $\hat h$ fixed, we first use the real scale and imaginary shift symmetry of the differential equation to write a more general $G$ with two real parameters $a$ and $b$.
\begin{align}
G &= - a \frac{i}{4} \frac{\cosh(w+\bar w)}{\cosh(2\bar w)} + \frac{i}{4} + i b
\end{align}
Next we construct $H$ for an arbitrary value of $c$ using $H = G/(1-i c G)$ with the requirement that $H$ satisfies the same boundary conditions as the $c=0$ case so that $H = 0$ when $y = 0$ and $H=-i$ when $y=\pi/2$.  This leads to the restriction $a = -2/(1+c)$ and $b=-(3+c)/4(1+c)$ with $c$ left as a parameter.  Again, for this choice one can check that $H$ satisfies the constraint \eqref{finalrangeconst}.
The corresponding metric factors as a function of $c$ are given by
\begin{align}
f_1^6 &= L^6 \frac{8 \cosh^2(2x)}{(1-c)^6(-1-c)} \Big[-1-c-(3+c)\cos(2y)+2 \cosh(4x)\Big] & \cr
& \qquad \qquad \Big[2-(3+c)\cos(2y)-(1+c)\cosh(4x)\Big] & \cr
(2 \rho)^6 &= L^6 \frac{[-1-c-(3+c)\cos(2y)+2\cosh(4x)][2-(3+c)\cos(2y)-(1+c)\cosh(4x)]}{(1+c)^4\cosh^4(2x)} & \cr
f_2^6 &= L^6 \frac{64 \cosh^2(2x)[2-(3+c)\cos(2y)-(1+c)\cosh(4x)] \cos^6(y)}{(1+c)^4 [1+c+(3+c)\cos(2y)-2 \cosh(4x)]^2} &\cr
f_3^6 &= L^6 \frac{8 \cosh^2(2x) [-1-c-(3+c)\cos(2y)+2 \cosh(4x)]\sin^6(y)}{(-1-c)[-2+(3+c)\cos(2y)+(1+c)\cosh(4x)]^2} &
\end{align}
One may readily check that the bulk geometry is regular when $c$ is in the range $-\infty < c < -1$.  In particular, the $S^3$'s cap off smoothly at the boundary of $\Sigma$.  The geometry contains singularities when $c$ is outside of this range, as can be seen by noting that $f_1$ vanishes for some value of $x$ and $y$.

There are more solutions one can construct.  For example using the same methods as above we can deform the solutions of \cite{D'Hoker:2008qm} or the Janus solutions of \cite{D'Hoker:2009gg}.  It is also possible that new types of solutions exist, which carry both M2 and M5 brane charges when $c$ is away from the special values $c \in \{-3,0,3\}$.  Another interesting possibility is to look for solutions which have $AdS_3$ asymptotics.  We leave the constructions of these solutions, as well as their implications for AdS/CFT to future work.

%%%%%%%%%%%%%%%%%%%%%%%%%%%%%%%%%%%%%%%%%%%%%
%%%%%%%%%%%%%%%%%%%%%%%%%%%%%%%%%%%%%%%%%%%%%
\section*{Acknowledgments}
The authors would like to thank E.~D'Hoker and M.~Gutperle for the fruitful collaborations leading to this work and E.~D'Hoker for useful comments on an earlier draft.
D.~K. was supported by KU Leuven grant OT/11/063.
J.~E. is supported by the FWO - Vlaanderen, Project No. G.0651.11, and by the ``Federal Office for Scientific, Technical and Cultural Affairs through the Inter-University Attraction Poles Programme,''  Belgian Science Policy P6/11-P.  This work is also supported by the European Science Foundation Holograv
Network.

\appendix

%\section{Derivation of Physical Fields}\label{sectionfields}
%\setcounter{equation}{0}

%In this section, we work out the expressions for the metric factors and fluxes in terms of the quantities $G$ and $h$.

\section{Metric factors}\label{sec:metric}
\setcounter{equation}{0}

Our first goal is to find the expressions for the metric factors in terms of $G$ and $h$. It turns out to be technically difficult to directly express $\alpha,\,\beta$ in terms of $G,\,\bar G$, so we first express $\alpha,\,\beta$ in terms of trigonometric functions of $\f,\,\bar \f$ which can be easily expressed in terms of $G,\,\bar G$. We define a new variable $k$ which is simply related to $\kappa$
\begin{align}\label{kkappa}
\bar k&=\rho(\ba\a^3-\bb\b^3)\\
\bar\k&=\bar k-c\rho\a\b(\ba\b-\a\bb)\no
\end{align}
Furthermore, consider the algebraic constraint given in the first equation of (\ref{kappaconstraint2}). The second term $\as\bs(\abs\bs-\as\bbs)$ is just $iG$. We then introduce $F$ and rewrite \eqref{kappaconstraint2} as $1=F-(c G/i)$ where
\begin{align}\label{k}
F&=1-i c G=\abs\as^3-\bbs\bs^3\no\\
\bar k&=\bar \kappa F=\bar\k(1-i c G)
\end{align}
Note that $F$ and $k$ are just $1$ and $\kappa$ for the special case of $c=0$.  We define $\f$, $\bar\f$ so that the first equation in (\ref{kkappa}) is automatic\footnote{This is similar to the definition used in \cite{D'Hoker:2008wc}, however here the function $k$ appears instead of the holomorphic function $\kappa$.}
\begin{align}\label{fidef}
&\sinh^2(\f)=\frac{\rho}{ k}\b\bb^3 &\qquad\qquad   \cosh^2(\f)=\frac{\rho}{ k}\a\ba^3
%&\cosh(2\f)={\rho \over k}(\ba\a^3+\bb\b^3) &\qquad\qquad \sinh(2\f)^2=4{\rho^2 \over k^2}\a\ba^3\b\bb^3
\end{align}
The complex conjugate expressions are also implied as usual.
The convenience of $\f$ is that these expressions are easily invertible and can be solved to obtain the following
\begin{align}\label{alphafi}
\a=\rho^{-{1\over 4}}\left({\bar k^3\over k}\right)^{1\over 8}\left({\cosh^3\bar\f\over\cosh\f}\right)^{1\over 4}\qquad\qquad
\b=\rho^{-{1\over 4}}\left({\bar k^3\over k}\right)^{1\over 8}\left({\sinh^3\bar\f\over\sinh\f}\right)^{1\over 4}
\end{align}

Now we express $G$ in terms of $\f$ starting from (\ref{Galphabeta}) and using the above equations to eliminate $\alpha$ and $\beta$.  The resulting expression for $G$ is
\begin{align}
G=-iF\sinh(\f-\bar\f)\left({\sinh(2\bar \f)\over\sinh(2\f)}\right)^{1\over 2}
\end{align}
From the above and its conjugate we obtain
\begin{align}
\sqrt{-H\bar H}=\sqrt{-G\bar G\over F\bar F}=\sinh(\f-\bar\f)\label{sinhfiimaginary} \\
{H\over \bar H}={G \bar F\over \bar G F}={\sinh(2\bar\f)\over\sinh(2\f)}\label{sinh2firatio}
\end{align}
where we have defined $H=G/F$.
For completeness, we also provide the following easily derivable but useful formulas
\begin{align}\label{sinh2fi}
\cosh(\f-\bar\f)=\sqrt{1-|H|^2}\no\\
\sinh(2(\f-\bar\f))=2\sqrt{|H|^2(|H|^2-1)}
\end{align}

Note that since $\f-\bar\f$ is pure imaginary then $|\cosh(\f-\bar\f)|\leq 1$ which implies $|H| \leq 1$.  Since $G$ and $H$ are related by $H = G/F$ with $F$ given by \eqref{k}, or equivalently
\begin{align}
\label{Hdef}
H = \frac{G}{1-i c G}
\end{align}
the restriction on the range of $H$ implies a restriction on the range of $G$.  This is further discussed in section \ref{subsec:ranges}.

We will need the expression for $\sinh(2\f)$, which requires quite a bit of algebra using the expressions \eqref{sinhfiimaginary} and \eqref{sinh2firatio}. %The derivation is in (\ref{sinhappendix})***add a bit more info here***
\begin{align}\label{sinh2f}
\sinh^2(2\f)={4\bar H^2(|H|^4-|H|^2)\over (H-\bar H)^2+4|H|^4}
\end{align}
Finally, in the derivation of the currents, we will need $\sinh(\f+\bar\f)$ which can be obtained from \eqref{sinhfiimaginary}, \eqref{sinh2fi} and \eqref{sinh2f}. Once again, branch cut choices are a difficulty and the sign between the two roots which appears in the expression below is chosen to be negative since $\sinh(\f+\bar\f)$ vanishes for imaginary $\f$ or equivalently imaginary $H$.
\begin{align}\label{sinhfr}
\sinh(\f+\bar\f)&=\sqrt{-|H|^2(1+\sinh^2(2\bar\f))}-\sqrt{(1-|H|^2)\sinh^2(2\bar\f)}\no\\
&={|H|(H+\bar H)\over\sqrt{-(H-\bar H)^2-4|H|^4}}
\end{align}

We can now obtain $\rho$ from the first line of (\ref{hmetrics}), (\ref{Mmetric}), and (\ref{alphafi}). Note that in (\ref{Mmetric}), the factor of $(|\kappa|/\rho)^{3/2}$ in the denominator is exactly what is needed to write $M$ entirely in terms of $\f$ and $\bar\f$ using  (\ref{alphafi}). Then $\rho$ is obtained by plugging $M(\f,\bar\f)$ into  the first line of (\ref{hmetrics}).
\begin{align}\label{rhofi}
\rho^{3\over 2}={-  |k|^{3\over 2} \over \sqrt 2 c_3 h}{\sinh(2(\f-\bar\f))\over|\sinh(2\f)|^{1\over 2}}
\end{align}
Using (\ref{k}, \ref{sinh2fi}, \ref{sinh2f}) to eliminate $\f$ in terms of $H$ we obtain
\begin{align}\label{rhoHFkappa}
\rho^6={|\kappa|^6|F|^6\over  c_3^4 h^4} (|H|^2-1)\Big((H-\bar H)^2+4|H|^4\Big)
\end{align}
Note that the first parenthesis is always negative since as we already argued $|H|\leq 1$. The second parenthesis is also always negative. To see the latter note that we can follow the logic of \cite{D'Hoker:2008wc} once again and say that
\begin{align}
(H-\bar H)^2+4|H|^4= -4\sin^2(\mu)\sin^2\theta+4    \sin^4(\mu) \no\\
i\mu=\f-\bar \f \qquad \qquad
e^{-i\theta}=\left({\sinh(2\bar \f)\over\sinh(2\f)}\right)^{1\over 2} \no
\end{align}
From the above definitions of $\theta$ and $\mu$ in terms of $\f$ and $\bar\f$ it can be shown that \linebreak[4]$ \tan\theta\tanh(\f+\bar\f)=\tan\mu $.  Using also that $|\tanh(\f+\bar\f)|\leq 1$, the former implies that $|\tan\mu|\leq|\tan\theta|$ which is equivalent to $|\sin\mu|\leq|\sin\theta|$, implying that
\begin{align}
(H-\bar H)^2+4|H|^4 \leq 0
\end{align}

Now we find $f_1$ by plugging (\ref{alphafi}) into (\ref{metric})
\begin{align}\label{f1start}
f_1={1\over c_1}(|\a|^2+|\b|^2)=\left({|F\k|\over c_1^2\rho}\right)^{1\over 2}(|\cosh\f|+|\sinh\f|)
\end{align}
By using the following trig identities
\begin{align}\label{coshplussinhtrig}
|\cosh\f|+|\sinh\f|&=\left(|\sinh(2\f)|\pm\sqrt{|\sinh(2\f)|^2+\cosh^2(\f-\bar\f)}\right)^{1\over 2}\\
|\cosh\f|-|\sinh\f|&=\left(-|\sinh(2\f)|\pm\sqrt{|\sinh(2\f)|^2+\cosh^2(\f-\bar\f)}\right)^{1\over 2}
\end{align}
and plugging in (\ref{sinh2fi}, \ref{sinh2f}) we obtain
\begin{align}\label{coshsinhH}
|\cosh\f|+|\sinh\f|&=\left({(|H|^2-1)\over (H-\bar H)^2+4|H|^4}\right)^{1\over 4}(2|H|^2\pm i(H-\bar H))^{1\over 2}\\
|\cosh\f|-|\sinh\f|&=\left({(|H|^2-1)\over (H-\bar H)^2+4|H|^4}\right)^{1\over 4}(-2|H|^2\pm i(H-\bar H))^{1\over 2}
\end{align}
Finally plugging (\ref{rhoHFkappa}, \ref{coshsinhH}) into (\ref{f1start}) we get
\begin{align}\label{f1Hh}
f_1^6={c_3^2 \over c_1^6}h^2(|H|^2-1){\big(2|H|^2\pm i(H-\bar H)\big)^3\over ((H-\bar H)^2+4|H|^4)^2}
\end{align}
The denominator is always positive. The first parenthesis in the numerator is always negative as has been already discussed and $h^2$ is also always negative since $h$ is an imaginary function. The sign in the second parenthesis depends on whether the imaginary part of $H$ is positive or negative and must be chosen such that the paranthesis overall is positive, in other words, such that the two terms in the parenthesis add. The reason for this is that this choice of sign exactly picks out $|\cosh\f|+|\sinh\f|$ over $|\cosh\f|-|\sinh\f|$ which picks out $f_1$ as opposed to $f_2$. Also note that although there are no roots appearing in the expression above, square roots appear during the derivation so there is an overall choice of sign in the above expression. The positive sign is chosen to give a positive expression.

Similarly, we also have
\begin{align}\label{f2start}
f_2={-1\over c_2}(|\a|^2-|\b|^2)=-\left({|F\k|\over c_2^2\rho}\right)^{1\over 2}(|\cosh\f|-|\sinh\f|)
\end{align}
Plugging in (\ref{rhoHFkappa}, \ref{coshsinhH}) we get
\begin{align}\label{f2Hh}
f_2^6=- {c_3^2 \over c_2^6}h^2(|H|^2-1){\big(2|H|^2\mp i(H-\bar H)\big)^3\over ((H-\bar H)^2+4|H|^4)^2}
\end{align}
This expression is very similar to above, but here the sign choice in the second parenthesis of the numerator is such that the two terms subtract, which is the sign choice corresponding to $|\cosh\f|-|\sinh\f|$. As we have argued above $|\textrm{Im}(H)|\geq |H|^2$ so this parenthesis is negative and the overall expression is positive.

Finally we come to the last metric factor
\begin{align}\label{f3start}
f_3={-i\over c_3}(\ba\b-\bb\a)={i\over c_3}\left({|\k F|\over\rho}\right)^{1\over 2}{\sinh(\f-\bar\f)\over\sqrt{1\over 2}|\sinh(2\f)|^{1\over 2}}
\end{align}
Plugging in (\ref{sinhfiimaginary}, \ref{sinh2f}, \ref{rhoHFkappa}) we get
\begin{align}\label{f3Hh}
f_3^6={h^2\over c_3^4}{((H-\bar H)^2+4|H|^4)\over (|H|^2-1)^2}
\end{align}

Note that the sign choices inside the numerators of $f_1$ and $f_2$ are correlated not only within the numerator but also between $f_1$ and $f_2$ themselves.  The sign is chosen such that $\pm i(H-\bar H)=|H-\bar H|$. This implies that we can get rid of the sign choice in the above expressions which are in terms of $H$ and write them with a unique sign using the absolute value function.  In the next section, we shall see that requiring the fluxes to satisfy the Bianchi identity selects the branch with $i(H-\bar H)=|H-\bar H|$, which means that $\textrm{Im}(H) \leq 0$.

Taking $c=0$ and $G=H$ in the final expressions for the metric factors \eqref{rhoHFkappa}, \eqref{f1Hh}, \eqref{f2Hh} and \eqref{f3Hh}, we recover the expressions for the metric factors obtained in section 5 of \cite{D'Hoker:2008wc}. Furthermore, if we take $c=-3$ and $G\rightarrow -(1/4)G+(i/4)$, we recover the metric factor expressions in section 7 of \cite{D'Hoker:2008wc}.  As discussed in detail in section \ref{subsec:compdisc}, one needs to perform a linear transformation on $G$ in order to take into account the difference in $G$ functions in \cite{D'Hoker:2008wc} and our paper.

\section{Field strength components}\label{sec:fluxes}
\setcounter{equation}{0}

Our next goal is to compute the field strength components, $g_{iz}$.  We will first compute $g_{1z}$ and $g_{2z}$.  To do so, we use the first two equations of \eqref{omegadiff} to express $g_{1z}$ and $g_{2z}$ in terms of $\bar \omega_{\alpha}$, $\bar \omega_{\beta}$, $\alpha_s$, $\beta_s$ and their complex conjugates.  Next we use the definitions of $\omega_{\alpha}$ and $\omega_{\beta}$ given above \eqref{omegadiff} and then \eqref{fidef} to express $\bar \omega_{\alpha}$ and $\bar \omega_{\beta}$ in terms of $F$ and $\bar \f$.  Finally, we eliminate the remaining $\alpha_s$ and $\beta_s$, by first rewriting them in terms of $\alpha$ and $\beta$ using \eqref{scaledvars}, then the relations between $\alpha$, $\beta$ and $\f$ given in \eqref{alphafi} and finally we use \eqref{k} to write $k$ in terms of $F$, with the overall $\kappa$ dependence now dropping out.  Putting everything together, we obtain
\begin{align}
\begin{split}
\label{gintermed}
\rho g_{1z}&={2^{{1\over 2}}|\sinh(2\f)|^{{1\over 2}}\over \sinh(\f+\bar\f)}(\p_w\log(F\cosh^2\bar\f)|\cosh\f|+\p_w\log(F\sinh^2\bar\f)|\sinh\f|) \cr
\rho g_{2z}&={2^{{1\over 2}}|\sinh(2\f)|^{{1\over 2}}\over \sinh(\f+\bar\f)}(-\p_w\log(F\cosh^2\bar\f)|\cosh\f|+\p_w\log(F\sinh^2\bar\f)|\sinh\f|)
\end{split}
\end{align}
The various quantities appearing in the above formula are given as follows: $|\cosh\f|$ and $|\sinh\f|$ can be computed using \eqref{coshsinhH},
$\sinh(\f+\bar\f)$  using \eqref{sinhfr} and $|\sinh(2\f)|$ using \eqref{sinh2f}.  To get $\cosh^2 \bar \f$ and $\sinh^2 \bar \f$, we use the double angle formulas to express them in terms of $\cosh(2\f)$, which in turn can be expressed as a root of $1+ \sinh^2(2 \f)$ with $\sinh(2 \f)$ given by \eqref{sinh2f}.

We continue the calculation by switching to computing $\p_w b_i=-\rho f_i^3 g_{iz}/2$, which are the quantities appearing directly in the Bianchi identities and equations of motion.\footnote{The factor of a half comes from our conventions for which $g_z = g_8-i g_9$ and $\p_w = (\p_x  - i \p_y)/2$.} We make use of equations \eqref{f1Hh} and \eqref{f2Hh} for the metric factors.  It is also necessary to remember that $F$ can be expressed in terms of $H$ by $F=(1+ i c H)^{-1}$.  In deriving the final formulas,  the following algebraic identity is frequently used
\begin{align}
\Big(2 H\bar H\pm i(H-\bar H)\Big)\Big(2 H\bar H\mp i(H-\bar H)\Big)=4 H^2\bar H^2+(H-\bar H)^2
\end{align}
It is important to note that in the expressions for $\cosh^2 \bar \f$ and $\sinh^2 \bar \f$, a branch cut choice must be made and can be
be determined by demanding that the Bianchi identities are satisfied.  This is also the sign choice for which the expressions reduce to the ones in \cite{D'Hoker:2008wc} for $c=0$.  Parameterizing, the branch cut choice for $\p_w b_1$ by introducing $\tilde \nu = \pm 1$, we have
\begin{align}
\p_w b_1 =&  i c_3 h(1-H \bar H)\Bigg(c_1^3(H+\bar H)\big((H-\bar H)^2+4H^2\bar H^2 \big)   \Bigg)^{-1}
\bigg\{-{4 c \p_w H(2 H\bar H+ i\nu(H-\bar H))^2 \over -i+c H}\no\\
&+\Big(H\bar H(H\bar H-1)(\bar H-H+2 H^2 \bar H)\Big)^{-1} \Big(\p_w \bar H H(H+\bar H)+\p_w H\bar H(H-3\bar H+4H\bar H^2)\Big)\no\\
&\Big[-2i\tilde \nu(\bar H-H+2H^2\bar H)(2H \bar H+i\nu(H-\bar H))
\cr&
+{2(\bar H-H+2H^2\bar H)^2(2 H \bar H+i\nu(H-\bar H))^2\over (H-\bar H)^2+4H^2\bar H^2}\Big] \bigg\}
\end{align}
We have also introduced $\nu=\pm 1$, which is given by the sign of $\textrm{Im}(-H)$ and parameterizes the sign choice in the metric factors.
Demanding the Binachi identity to hold, one finds that $\tilde \nu=-\nu$.  The same choice arises in the calculation of $\p_w b_2$ and the sign must be chosen in the same way.
There is another sign ambiguity, since $f_1^3$ and $f_2^3$ are determined only up to a sign, we parameterize these sign choices by including overall factors $\nu_1 = \pm 1$ and $\nu_2 = \pm 1$.  The final expressions for $\p_w b_1$ and $\p_w b_2$ are given in \eqref{g1Hfinal} and \eqref{g2Hfinal}.

To compute $g_{3z}$, we first use the second equation in \eqref{algebraic2} to express $g_{3z}$ in terms of $g_{1z}$, $g_{2z}$ and $\alpha$ and $\beta$.  The remaining $\alpha$ and $\beta$ dependence can be expressed in terms of $\f$ using \eqref{alphafi}.  The result is
\begin{align}
\frac{\alpha^2 \pm \beta^2}{\alpha \beta} &= \sqrt{2} \frac{[ \sinh(\f-\bar \f)+\sinh(\f+\bar \f) \cosh(2 \bar \f) \pm \sinh^\frac{3}{2}(2 \bar \f) \sinh^\frac{1}{2}(2\f)]^\frac{1}{2}}{\sinh^\frac{3}{4}(2 \bar \f) \sinh^\frac{1}{4}(2 \f)}
\end{align}
Most of these quantities have already appeared in \eqref{gintermed}.  The only new term is $\sinh(\f- \bar \f)$ which is given by \eqref{sinhfiimaginary}.  Again one has to make choices for the signs of branch cuts.  The intermediate steps are messy and we do not include them here.  As before demanding the Bianchi identity to hold for $g_{3z}$ restricts the possible branch choices.  In this case however, one finds that the Bianchi identity for $g_{3z}$ holds only for the branch with $\nu = 1$ or equivalently $\textrm{Im}(H)<0$, except in the special case that $c=0$. This is discussed further in section \ref{subsec:ranges}.  The final result for $g_{3z}$ is given in \eqref{g3Hfinal}.

\section{Bianchi identities and equations of motion}\label{sectionbianchi}
\setcounter{equation}{0}

Several theorems exist guaranteeing that under certain conditions, solutions of the BPS equations are automatically supergravity solutions \cite{Gauntlett:2002fz, Gauntlett:2003wb}.  More specifically, one needs to check the Bianchi identities and Maxwell equations, however the Einstein equations are then either automatic or automatic up to a single equation depending on whether the Killing spinors yield time-like or null Killing vectors.

In deriving the explicit solutions for the metric factors and fluxes, we have to make sign choices for various branch cuts.  These sign choices must be made consistently with the BPS equations, which requires checking if the BPS equations enforce additional constraints on the sign choices.  The simpler alternative which we follow here is to check the equations of motion.  We note that the equations of motion can still leave sign choices for the fluxes undetermined, with some choices breaking supersymmetry, and one must still check the BPS equations \cite{Duff:1983nu,Duff:1986hr}.

In \cite{D'Hoker:2008wc}, the equations of motion and metric factors were checked for the special case $c = 0$ and we use their methods here.
We first make a change of notation and exchange the imaginary harmonic function $h$ for a real harmonic function $\hat h$ with the identification $\hat h=- i h$.  Next, we note that the metric factors and field strength components define a solution only for $G$ which satisfies its differential equation (\ref{Geqn}).  In order to implement this constraint in general, we make a conformal transformation on $\Sigma$ to coordinates defined by $\hat h$ and its dual harmonic function, $\tilde h$.
\begin{align}
&u=\hat h+i \tilde h&   &\bar u=\hat h-i\tilde h& \no\\
&\p_u(\hat h-i\tilde h)=0& &\p_{\bar u}(\hat h+i\tilde h)=0& \no\\
&2\p_u =  \p_{\hat h} - i \p_{\tilde h}&   &2 \p_{\bar u} =  \p_{\hat h} + i \p_{\tilde h}&
\end{align}
Equation (\ref{Geqn}) can be decomposed into its real and imaginary parts.
\begin{align}
&G(\hat h,\tilde h)  = G_R (\hat h,\tilde h) + i G_I (\hat h,\tilde h)\cr
&\p_{\hat h} G_R + \p_{\tilde h} G_I  =  { G_{R} \over \hat h}\cr
&\p_{\hat h} G_I - \p_{\tilde h} G_R  =  0
\end{align}
The second equation above demands that $G_R$ and $G_I$ are the derivatives of a single real function $\phi$ and then the first equation is rewritten as a second order equation on $\phi$.
\begin{align}
&G_R   =  \p_{\hat h}  ( \phi   )     \qquad \qquad \qquad G_I   =  \p_{\tilde h} ( \phi   )  \cr
& {1\over \hat h}\Big ( \p_{\tilde h}^2 + \p_{\hat h} ^2 - { 1 \over \hat h} \p_{\hat h}   \Big )\phi  =0   \label{phieq}
\end{align}
Finally, we use this equation to eliminate all derivatives in $\hat h$ which are of order two or higher, which implements the differential constraint.  To be specific, we eliminate $\p^3_{\hat h} G$, $\p^2_{\hat h} G$ and $\p^2_{\hat h} \p_{\tilde h} G$.

We begin with the Bianchi identity.
\begin{align}\label{bianchi}
&dF=0& &\Rightarrow&  &\p_w [\p_{\bar w} b_i] - c.c. =0& &i=\{1,2,3\} &
\end{align}
Using the expressions for $\p_{\bar w} b_i$ given in \eqref{g1Hfinal}, \eqref{g2Hfinal} and \eqref{g3Hfinal}, one finds that the Binachi identities are automatically satisfied after first expressing $G$ in terms of $\phi$ and using \eqref{phieq} to exclude all second and higher order $\hat h$ derivatives which appear.  Next we check the equation of motion of the field strength.
\bea
d * F + {1 \over 2} F \wedge F = 0
\eea
Using the anstaz given in section \ref{sectionsetup} and writing out in components we have\footnote{We note that \cite{D'Hoker:2008wc} contains a typo, where the factor of $1/4$ in the third term of each equation should be a $1/2$.}
\bea
\label{maxwelleqns}
0 &=&
\p_{\bar w} \p_w b_1
+ \half \left ( \p_{\bar w} b_1 \p_w \ln \bigg ( {f_2 f_3 \over f_1} \bigg ) ^3 + c.c. \right )
+ {i \over 2} \bigg( {f_1 \over f_2 f_3} \bigg)^3 (\p_w b_2 \p_{\bar w} b_3 - c.c.)
\no\\
0 &=&
\p_{\bar w} \p_w b_2
+ \half \left ( \p_{\bar w} b_2 \p_w \ln \bigg ( {f_1 f_3 \over f_2} \bigg ) ^3 + c.c. \right )
+ {i \over 2} \bigg( {f_2 \over f_1 f_3} \bigg)^3 (\p_w b_1 \p_{\bar w} b_3 - c.c.)
\no\\
0 &=&
\p_{\bar w} \p_w b_3
+ \half \left ( \p_{\bar w} b_3 \p_w \ln \bigg ( {f_1 f_2 \over f_3} \bigg ) ^3 + c.c. \right )
- {i \over 2} \bigg( {f_3 \over f_1 f_2} \bigg)^3 (\p_w b_1 \p_{\bar w} b_2 - c.c.)
\eea Again, one may check that these expressions are automatic after using \eqref{g1Hfinal}, \eqref{g2Hfinal} and \eqref{g3Hfinal} for the fluxes and \eqref{f1Hh}, \eqref{f2Hh} and \eqref{f3Hh} for the metric factors.  We remind the reader that the Bianchi identity for $g_{3z}$ selects the branch with $i(H-\bar H)\geq 0$.  We note that these equations are invariant under $b_1 \rightarrow - b_1$ along with a flip in sign of all of the metric factors so that $f_1 f_2 f_3 \rightarrow - f_1 f_2 f_3$.  Similar sign flips can be made for $b_2$ and $b_3$, thus the Maxwell equations determine the fluxes up to an overall sign.  We can determine the signs by considering the special cases $c = \{-3,0,3\}$ where we take the geometry to be either $AdS_4 \times S^7$ ($c=-3$) or $AdS_7 \times S^4$ ($c=0,3$).

Finally, the Einstein equations can be checked. These are equations (9.19) and (9.20) of \cite{D'Hoker:2008wc}. Some details of the derivation of these equations can be found there as well. The method for checking these equations is the same as with the Maxwell equations and Bianchi identities and requires replacing
second and third order derivatives using \ref{phieq}.

%%%%%%%%%%%%%%%%%%%%%%%%%%%%%%%%%%%%%%%%%%%%%%%%%%%%%%%%%%%%%%%%%%%%%%


\begin{thebibliography}{99}

%\cite{D'Hoker:2008ix}
\bibitem{D'Hoker:2008ix}
  E.~D'Hoker, J.~Estes, M.~Gutperle, D.~Krym and P.~Sorba,
  ``Half-BPS supergravity solutions and superalgebras,''
  JHEP {\bf 0812} (2008) 047
  [arXiv:0810.1484 [hep-th]].
  %%CITATION = ARXIV:0810.1484;%%
  %14 citations counted in INSPIRE as of 04 Mar 2013


%\cite{D'Hoker:2008wc}
\bibitem{D'Hoker:2008wc}
  E.~D'Hoker, J.~Estes, M.~Gutperle and D.~Krym,
  ``Exact Half-BPS Flux Solutions in M-theory. I: Local Solutions,''
  JHEP {\bf 0808} (2008) 028
  [arXiv:0806.0605 [hep-th]].
  %%CITATION = ARXIV:0806.0605;%%
  %26 citations counted in INSPIRE as of 04 Mar 2013


%\cite{Lunin:2007ab}
\bibitem{Lunin:2007ab}
  O.~Lunin,
  ``1/2-BPS states in M theory and defects in the dual CFTs,''
  JHEP {\bf 0710} (2007) 014
  [arXiv:0704.3442 [hep-th]].
  %%CITATION = ARXIV:0704.3442;%%
  %39 citations counted in INSPIRE as of 04 Mar 2013


%\cite{Maldacena:1997re}
\bibitem{Maldacena:1997re}
  J.~M.~Maldacena,
  ``The Large N limit of superconformal field theories and supergravity,''
  Adv.\ Theor.\ Math.\ Phys.\  {\bf 2} (1998) 231
  [hep-th/9711200].
  %%CITATION = HEP-TH/9711200;%%
  %8855 citations counted in INSPIRE as of 04 Mar 2013


%\cite{Howe:1997fb}
\bibitem{Howe:1997fb}
  P.~S.~Howe, E.~Sezgin and P.~C.~West,
  ``Covariant field equations of the M theory five-brane,''
  Phys.\ Lett.\ B {\bf 399} (1997) 49
  [hep-th/9702008].
  %%CITATION = HEP-TH/9702008;%%
  %198 citations counted in INSPIRE as of 04 Mar 2013


%\cite{Berkooz:1997cq}
\bibitem{Berkooz:1997cq}
  M.~Berkooz, M.~Rozali and N.~Seiberg,
  ``Matrix description of M theory on T**4 and T**5,''
  Phys.\ Lett.\ B {\bf 408} (1997) 105
  [hep-th/9704089].
  %%CITATION = HEP-TH/9704089;%%
  %209 citations counted in INSPIRE as of 04 Mar 2013


%\cite{Claus:1997cq}
\bibitem{Claus:1997cq}
  P.~Claus, R.~Kallosh and A.~Van Proeyen,
  ``M five-brane and superconformal (0,2) tensor multiplet in six-dimensions,''
  Nucl.\ Phys.\ B {\bf 518} (1998) 117
  [hep-th/9711161].
  %%CITATION = HEP-TH/9711161;%%
  %112 citations counted in INSPIRE as of 04 Mar 2013


%\cite{Aharony:1997an}
\bibitem{Aharony:1997an}
  O.~Aharony, M.~Berkooz and N.~Seiberg,
  ``Light cone description of (2,0) superconformal theories in six-dimensions,''
  Adv.\ Theor.\ Math.\ Phys.\  {\bf 2} (1998) 119
  [hep-th/9712117].
  %%CITATION = HEP-TH/9712117;%%
  %141 citations counted in INSPIRE as of 04 Mar 2013


%\cite{Douglas:2010iu}
\bibitem{Douglas:2010iu}
  M.~R.~Douglas,
  ``On D=5 super Yang-Mills theory and (2,0) theory,''
  JHEP {\bf 1102} (2011) 011
  [arXiv:1012.2880 [hep-th]].
  %%CITATION = ARXIV:1012.2880;%%
  %68 citations counted in INSPIRE as of 04 Mar 2013


%\cite{Lambert:2010iw}
\bibitem{Lambert:2010iw}
  N.~Lambert, C.~Papageorgakis and M.~Schmidt-Sommerfeld,
  ``M5-Branes, D4-Branes and Quantum 5D super-Yang-Mills,''
  JHEP {\bf 1101} (2011) 083
  [arXiv:1012.2882 [hep-th]].
  %%CITATION = ARXIV:1012.2882;%%
  %73 citations counted in INSPIRE as of 04 Mar 2013


%\cite{Kim:2011mv}
\bibitem{Kim:2011mv}
  H.~-C.~Kim, S.~Kim, E.~Koh, K.~Lee and S.~Lee,
  ``On instantons as Kaluza-Klein modes of M5-branes,''
  JHEP {\bf 1112} (2011) 031
  [arXiv:1110.2175 [hep-th]].
  %%CITATION = ARXIV:1110.2175;%%
  %24 citations counted in INSPIRE as of 04 Mar 2013


%\cite{Kim:2012av}
\bibitem{Kim:2012av}
  Y.~S.~Kim and M.~E.~Noz,
  ``Dirac Matrices and Feynman's Rest of the Universe,''
  Symmetry {\bf 4} (2012) 626
  [arXiv:1210.6251 [quant-ph]].
  %%CITATION = ARXIV:1210.6251;%%


%\cite{Jafferis:2012iv}
\bibitem{Jafferis:2012iv}
  D.~L.~Jafferis and S.~S.~Pufu,
  ``Exact results for five-dimensional superconformal field theories with gravity duals,''
  arXiv:1207.4359 [hep-th].
  %%CITATION = ARXIV:1207.4359;%%
  %19 citations counted in INSPIRE as of 04 Mar 2013


%\cite{Strominger:1995ac}
\bibitem{Strominger:1995ac}
  A.~Strominger,
  ``Open p-branes,''
  Phys.\ Lett.\ B {\bf 383} (1996) 44
  [hep-th/9512059].
  %%CITATION = HEP-TH/9512059;%%
  %460 citations counted in INSPIRE as of 04 Mar 2013


%\cite{Ganor:1996nf}
\bibitem{Ganor:1996nf}
  O.~J.~Ganor,
  ``Six-dimensional tensionless strings in the large N limit,''
  Nucl.\ Phys.\ B {\bf 489} (1997) 95
  [hep-th/9605201].
  %%CITATION = HEP-TH/9605201;%%
  %43 citations counted in INSPIRE as of 04 Mar 2013


%\cite{Howe:1997ue}
\bibitem{Howe:1997ue}
  P.~S.~Howe, N.~D.~Lambert and P.~C.~West,
  ``The Selfdual string soliton,''
  Nucl.\ Phys.\ B {\bf 515} (1998) 203
  [hep-th/9709014].
  %%CITATION = HEP-TH/9709014;%%
  %171 citations counted in INSPIRE as of 04 Mar 2013


%\cite{Saemann:2010cp}
\bibitem{Saemann:2010cp}
  C.~Saemann,
  ``Constructing Self-Dual Strings,''
  Commun.\ Math.\ Phys.\  {\bf 305} (2011) 513
  [arXiv:1007.3301 [hep-th]].
  %%CITATION = ARXIV:1007.3301;%%
  %17 citations counted in INSPIRE as of 04 Mar 2013


%\cite{D'Hoker:2008qm}
\bibitem{D'Hoker:2008qm}
  E.~D'Hoker, J.~Estes, M.~Gutperle and D.~Krym,
  ``Exact Half-BPS Flux Solutions in M-theory II: Global solutions asymptotic to AdS(7) x S**4,''
  JHEP {\bf 0812} (2008) 044
  [arXiv:0810.4647 [hep-th]].
  %%CITATION = ARXIV:0810.4647;%%
  %13 citations counted in INSPIRE as of 04 Mar 2013


%\cite{Bagger:2006sk}
\bibitem{Bagger:2006sk}
  J.~Bagger and N.~Lambert,
  ``Modeling Multiple M2's,''
  Phys.\ Rev.\ D {\bf 75} (2007) 045020
  [hep-th/0611108].
  %%CITATION = HEP-TH/0611108;%%
  %519 citations counted in INSPIRE as of 04 Mar 2013


%\cite{Bagger:2007jr}
\bibitem{Bagger:2007jr}
  J.~Bagger and N.~Lambert,
  ``Gauge symmetry and supersymmetry of multiple M2-branes,''
  Phys.\ Rev.\ D {\bf 77} (2008) 065008
  [arXiv:0711.0955 [hep-th]].
  %%CITATION = ARXIV:0711.0955;%%
  %632 citations counted in INSPIRE as of 04 Mar 2013


%\cite{Bagger:2007vi}
\bibitem{Bagger:2007vi}
  J.~Bagger and N.~Lambert,
  ``Comments on multiple M2-branes,''
  JHEP {\bf 0802} (2008) 105
  [arXiv:0712.3738 [hep-th]].
  %%CITATION = ARXIV:0712.3738;%%
  %444 citations counted in INSPIRE as of 04 Mar 2013


%\cite{Gustavsson:2007vu}
\bibitem{Gustavsson:2007vu}
  A.~Gustavsson,
  ``Algebraic structures on parallel M2-branes,''
  Nucl.\ Phys.\ B {\bf 811} (2009) 66
  [arXiv:0709.1260 [hep-th]].
  %%CITATION = ARXIV:0709.1260;%%
  %583 citations counted in INSPIRE as of 04 Mar 2013


%\cite{Aharony:2008ug}
\bibitem{Aharony:2008ug}
  O.~Aharony, O.~Bergman, D.~L.~Jafferis and J.~Maldacena,
  ``N=6 superconformal Chern-Simons-matter theories, M2-branes and their gravity duals,''
  JHEP {\bf 0810} (2008) 091
  [arXiv:0806.1218 [hep-th]].
  %%CITATION = ARXIV:0806.1218;%%
  %935 citations counted in INSPIRE as of 04 Mar 2013


%\cite{D'Hoker:2009gg}
\bibitem{D'Hoker:2009gg}
  E.~D'Hoker, J.~Estes, M.~Gutperle and D.~Krym,
  ``Janus solutions in M-theory,''
  JHEP {\bf 0906} (2009) 018
  [arXiv:0904.3313 [hep-th]].
  %%CITATION = ARXIV:0904.3313;%%
  %10 citations counted in INSPIRE as of 04 Mar 2013


%\cite{Bak:2003jk}
\bibitem{Bak:2003jk}
  D.~Bak, M.~Gutperle and S.~Hirano,
  ``A Dilatonic deformation of AdS(5) and its field theory dual,''
  JHEP {\bf 0305} (2003) 072
  [hep-th/0304129].
  %%CITATION = HEP-TH/0304129;%%
  %75 citations counted in INSPIRE as of 04 Mar 2013


%\cite{Clark:2005te}
\bibitem{Clark:2005te}
  A.~Clark and A.~Karch,
  ``Super Janus,''
  JHEP {\bf 0510} (2005) 094
  [hep-th/0506265].
  %%CITATION = HEP-TH/0506265;%%
  %28 citations counted in INSPIRE as of 04 Mar 2013


%\cite{D'Hoker:2006uu}
\bibitem{D'Hoker:2006uu}
  E.~D'Hoker, J.~Estes and M.~Gutperle,
  ``Ten-dimensional supersymmetric Janus solutions,''
  Nucl.\ Phys.\ B {\bf 757} (2006) 79
  [hep-th/0603012].
  %%CITATION = HEP-TH/0603012;%%
  %26 citations counted in INSPIRE as of 04 Mar 2013


%\cite{Clark:2004sb}
\bibitem{Clark:2004sb}
  A.~B.~Clark, D.~Z.~Freedman, A.~Karch and M.~Schnabl,
  ``The Dual of Janus ((<:)<->(:>)) an interface CFT,''
  Phys.\ Rev.\ D {\bf 71} (2005) 066003
  [hep-th/0407073].
  %%CITATION = HEP-TH/0407073;%%
  %47 citations counted in INSPIRE as of 04 Mar 2013


%\cite{D'Hoker:2006uv}
\bibitem{D'Hoker:2006uv}
  E.~D'Hoker, J.~Estes and M.~Gutperle,
  ``Interface Yang-Mills, supersymmetry, and Janus,''
  Nucl.\ Phys.\ B {\bf 753} (2006) 16
  [hep-th/0603013].
  %%CITATION = HEP-TH/0603013;%%
  %37 citations counted in INSPIRE as of 04 Mar 2013


%\cite{Gaiotto:2008sd}
\bibitem{Gaiotto:2008sd}
  D.~Gaiotto and E.~Witten,
  ``Janus Configurations, Chern-Simons Couplings, And The theta-Angle in N=4 Super Yang-Mills Theory,''
  JHEP {\bf 1006} (2010) 097
  [arXiv:0804.2907 [hep-th]].
  %%CITATION = ARXIV:0804.2907;%%
  %159 citations counted in INSPIRE as of 04 Mar 2013


%\cite{Honma:2008un}
\bibitem{Honma:2008un}
  Y.~Honma, S.~Iso, Y.~Sumitomo and S.~Zhang,
  ``Janus field theories from multiple M2 branes,''
  Phys.\ Rev.\ D {\bf 78} (2008) 025027
  [arXiv:0805.1895 [hep-th]].
  %%CITATION = ARXIV:0805.1895;%%
  %53 citations counted in INSPIRE as of 04 Mar 2013


%\cite{D'Hoker:2009my}
\bibitem{D'Hoker:2009my}
  E.~D'Hoker, J.~Estes, M.~Gutperle and D.~Krym,
  ``Exact Half-BPS Flux Solutions in M-theory III: Existence and rigidity of global solutions asymptotic to AdS(4) x S**7,''
  JHEP {\bf 0909} (2009) 067
  [arXiv:0906.0596 [hep-th]].
  %%CITATION = ARXIV:0906.0596;%%
  %14 citations counted in INSPIRE as of 04 Mar 2013


%\cite{Basu:2004ed}
\bibitem{Basu:2004ed}
  A.~Basu and J.~A.~Harvey,
  ``The M2-M5 brane system and a generalized Nahm's equation,''
  Nucl.\ Phys.\ B {\bf 713} (2005) 136
  [hep-th/0412310].
  %%CITATION = HEP-TH/0412310;%%
  %141 citations counted in INSPIRE as of 04 Mar 2013


%\cite{Jeon:2008bx}
\bibitem{Jeon:2008bx}
  I.~Jeon, J.~Kim, N.~Kim, S.~-W.~Kim and J.~-H.~Park,
  ``Classification of the BPS states in Bagger-Lambert Theory,''
  JHEP {\bf 0807} (2008) 056
  [arXiv:0805.3236 [hep-th]].
  %%CITATION = ARXIV:0805.3236;%%
  %57 citations counted in INSPIRE as of 04 Mar 2013


%\cite{Hanaki:2008cu}
\bibitem{Hanaki:2008cu}
  K.~Hanaki and H.~Lin,
  ``M2-M5 Systems in N=6 Chern-Simons Theory,''
  JHEP {\bf 0809} (2008) 067
  [arXiv:0807.2074 [hep-th]].
  %%CITATION = ARXIV:0807.2074;%%
  %57 citations counted in INSPIRE as of 04 Mar 2013


%\cite{Jeon:2008zj}
\bibitem{Jeon:2008zj}
  I.~Jeon, J.~Kim, B.~-H.~Lee, J.~-H.~Park and N.~Kim,
  ``M-brane bound states and the supersymmetry of BPS solutions in the Bagger-Lambert theory,''
  Int.\ J.\ Mod.\ Phys.\ A {\bf 24} (2009) 5779
  [arXiv:0809.0856 [hep-th]].
  %%CITATION = ARXIV:0809.0856;%%
  %21 citations counted in INSPIRE as of 04 Mar 2013


%\cite{Ammon:2009wc}
\bibitem{Ammon:2009wc}
  M.~Ammon, J.~Erdmenger, R.~Meyer, A.~O'Bannon and T.~Wrase,
  ``Adding Flavor to AdS(4)/CFT(3),''
  JHEP {\bf 0911} (2009) 125
  [arXiv:0909.3845 [hep-th]].
  %%CITATION = ARXIV:0909.3845;%%
  %17 citations counted in INSPIRE as of 04 Mar 2013


%\cite{Berman:2009xd}
\bibitem{Berman:2009xd}
  D.~S.~Berman, M.~J.~Perry, E.~Sezgin and D.~C.~Thompson,
  ``Boundary Conditions for Interacting Membranes,''
  JHEP {\bf 1004} (2010) 025
  [arXiv:0912.3504 [hep-th]].
  %%CITATION = ARXIV:0912.3504;%%
  %18 citations counted in INSPIRE as of 04 Mar 2013


%\cite{Fujimori:2010ec}
\bibitem{Fujimori:2010ec}
  T.~Fujimori, K.~Iwasaki, Y.~Kobayashi and S.~Sasaki,
  ``Classification of BPS Objects in N = 6 Chern-Simons Matter Theory,''
  JHEP {\bf 1010} (2010) 002
  [arXiv:1007.1588 [hep-th]].
  %%CITATION = ARXIV:1007.1588;%%
  %7 citations counted in INSPIRE as of 04 Mar 2013


%\cite{Fujita:2010pj}
\bibitem{Fujita:2010pj}
  M.~Fujita,
  ``M5-brane Defect and QHE in $AdS_4 \times N(1,1)/N=3$ SCFT,''
  Phys.\ Rev.\ D {\bf 83} (2011) 105016
  [arXiv:1011.0154 [hep-th]].
  %%CITATION = ARXIV:1011.0154;%%
  %9 citations counted in INSPIRE as of 04 Mar 2013

%\cite{Colgain:2012wv}
\bibitem{Colgain:2012wv}
  E.~OColgain,
  ``Beyond LLM in M-theory,''
  arXiv:1208.5979 [hep-th].
  %%CITATION = ARXIV:1208.5979;%%

%\cite{Gustavsson:2010ep}
\bibitem{Gustavsson:2010ep}
  A.~Gustavsson,
  ``Five-Dimensional Super Yang-Mills Theory from ABJM Theory,''
  JHEP {\bf 1103} (2011) 144
  [arXiv:1012.5917 [hep-th]].
  %%CITATION = ARXIV:1012.5917;%%
  %5 citations counted in INSPIRE as of 04 Mar 2013


%\cite{Terashima:2010ji}
\bibitem{Terashima:2010ji}
  S.~Terashima and F.~Yagi,
  ``On Effective Action of Multiple M5-branes and ABJM Action,''
  JHEP {\bf 1103} (2011) 036
  [arXiv:1012.3961 [hep-th]].
  %%CITATION = ARXIV:1012.3961;%%
  %11 citations counted in INSPIRE as of 04 Mar 2013


%\cite{Lambert:2011eg}
\bibitem{Lambert:2011eg}
  N.~Lambert, H.~Nastase and C.~Papageorgakis,
  ``5D Yang-Mills instantons from ABJM Monopoles,''
  Phys.\ Rev.\ D {\bf 85} (2012) 066002
  [arXiv:1111.5619 [hep-th]].
  %%CITATION = ARXIV:1111.5619;%%
  %13 citations counted in INSPIRE as of 04 Mar 2013


%\cite{Gomis:2008vc}
\bibitem{Gomis:2008vc}
  J.~Gomis, D.~Rodriguez-Gomez, M.~Van Raamsdonk and H.~Verlinde,
  ``A Massive Study of M2-brane Proposals,''
  JHEP {\bf 0809} (2008) 113
  [arXiv:0807.1074 [hep-th]].
  %%CITATION = ARXIV:0807.1074;%%
  %108 citations counted in INSPIRE as of 04 Mar 2013


%\cite{Hosomichi:2008jb}
\bibitem{Hosomichi:2008jb}
  K.~Hosomichi, K.~-M.~Lee, S.~Lee, S.~Lee and J.~Park,
  ``N=5,6 Superconformal Chern-Simons Theories and M2-branes on Orbifolds,''
  JHEP {\bf 0809} (2008) 002
  [arXiv:0806.4977 [hep-th]].
  %%CITATION = ARXIV:0806.4977;%%
  %198 citations counted in INSPIRE as of 04 Mar 2013


%\cite{Gauntlett:2002fz}
\bibitem{Gauntlett:2002fz}
  J.~P.~Gauntlett and S.~Pakis,
  ``The Geometry of D = 11 killing spinors,''
  JHEP {\bf 0304} (2003) 039
  [hep-th/0212008].
  %%CITATION = HEP-TH/0212008;%%
  %206 citations counted in INSPIRE as of 04 Mar 2013


%\cite{Gauntlett:2003wb}
\bibitem{Gauntlett:2003wb}
  J.~P.~Gauntlett, J.~B.~Gutowski and S.~Pakis,
  ``The Geometry of D = 11 null Killing spinors,''
  JHEP {\bf 0312} (2003) 049
  [hep-th/0311112].
  %%CITATION = HEP-TH/0311112;%%
  %91 citations counted in INSPIRE as of 04 Mar 2013


%\cite{Duff:1983nu}
\bibitem{Duff:1983nu}
  M.~J.~Duff, B.~E.~W.~Nilsson and C.~N.~Pope,
  ``Spontaneous Supersymmetry Breaking By The Squashed Seven Sphere,''
  Phys.\ Rev.\ Lett.\  {\bf 50} (1983) 2043.
  %%CITATION = PRLTA,50,2043;%%
  %102 citations counted in INSPIRE as of 04 Mar 2013


%\cite{Duff:1986hr}
\bibitem{Duff:1986hr}
  M.~J.~Duff, B.~E.~W.~Nilsson and C.~N.~Pope,
  ``Kaluza-Klein Supergravity,''
  Phys.\ Rept.\  {\bf 130} (1986) 1.
  %%CITATION = PRPLC,130,1;%%
  %642 citations counted in INSPIRE as of 04 Mar 2013

  \end{thebibliography}
\end{document}